\documentclass[fleqn,usenatbib,a4]{mnras}

\usepackage{newtxtext,newtxmath}
\usepackage[T1]{fontenc}

\DeclareRobustCommand{\VAN}[3]{#2}
\let\VANthebibliography\thebibliography
\def\thebibliography{\DeclareRobustCommand{\VAN}[3]{##3}\VANthebibliography}

\usepackage{savesym}
\savesymbol{tablenum}

\usepackage{graphicx}%
\usepackage{multirow}%
\usepackage{amsmath}%
\usepackage{xcolor}%
\usepackage{xfrac}
\usepackage{mismath}
\usepackage{float}

\usepackage{chemformula}
\usepackage{hyperref}
\usepackage[separate-uncertainty=true]{siunitx}
\usepackage{lipsum}

\newcommand{\dd}{\mathop{}\!\mathrm{d}}

\definecolor{revisegreen}{RGB}{45,105,15}
\definecolor{revisepink}{RGB}{153,0,255}

\DeclareSIUnit\bar{bar}
\DeclareSIUnit\AU{AU}
\DeclareSIUnit\dex{dex}
\DeclareSIUnit\erg{erg}
\DeclareSIUnit\day{day}
\DeclareSIUnit\year{yr}
\DeclareSIUnit\soli{So}
\DeclareSIUnit\amu{\gram\per\mole}
\DeclareSIUnit\WPMS{\watt\per\meter\squared}
\DeclareSIUnit\um{\micro\meter}
\DeclareSIUnit\NWPKG{\nano\watt\per\kilo\gram}
\DeclareSIUnit\GPCC{\gram\per\centi\meter\cubed}
\DeclareSIUnit\GPa{\giga\pascal}
\DeclareSIUnit\cmsps{\centi\meter\squared\per\second}
\DeclareSIUnit\mpss{\meter\per\second\squared}

\newcommand{\PiMen}{{$\pi$\,Men\,}}
\newcommand{\Kzz}{{$K_\mathrm{zz}$}}

\title[Beyond the mass-radius plane]{Beyond the mass-radius plane: Integrated radiative-convective and interior structure simulations of the exoplanet continuum}

\author[H. Nicholls et al.]{
Harrison Nicholls,$^{1,2}$\thanks{E-mail: harrison.nicholls@ast.cam.ac.uk}
Oliver Shorttle,$^{1}$
Tim Lichtenberg,$^{3}$
and
Flavia Pascal$^{1,3}$
\\
$^{1}$Institute of Astronomy, University of Cambridge, United Kingdom\\
$^{2}$Atmospheric Oceanic and Planetary Physics, University of Oxford, United Kingdom\\
$^{3}$Kapteyn Astronomical Institute, University of Groningen, The Netherlands
}

\date{Accepted XXX. Received YYY; in original form ZZZ}
\pubyear{\the\year{}}
\begin{document}
\label{firstpage}
\pagerange{\pageref{firstpage}--\pageref{lastpage}}
\maketitle

\begin{abstract}
Static structure models, which map mass-radius constraints to bulk planet composition, are frequently used to categorise exoplanets due to their computational efficiency and the high-level insight they offer into planetary properties. However, static structure models typically have simplified atmospheric treatments, which may introduce systematic biases when interpreting the structures -- and therefore the climates -- of sub-Neptunes and super-Earths.
We present a framework for recovering exoplanet properties using static structure models that accounts for necessary physical-chemical complexity in their atmospheres. We produce a comprehensive library of 504,000 exoplanet simulations that unify deep planetary interior structure with radiative-convective-chemical climate calculations. From these models we demonstrate that a planet's envelope mass fraction -- a critical parameter to infer -- is frequently degenerate with its instellation flux and atmospheric metallicity, and sensitive to the treatment of gravitational acceleration at the mbar level. Such uncertainties have significant implications for inferring planetary processes, as our modelling shows that habitable-zone sub-Neptunes readily host supercritical surfaces or deep magma oceans, despite their temperate irradiation regime. To marginalise over these uncertainties, we introduce a Bayesian retrieval tool that uses our library of self-consistent models. By applying this Bayesian approach to case-studies of \PiMen\,c and TOI-421\,b, we show that robust physical interpretations are achievable through whole-planet mass-radius retrievals. While new data from JWST, Ariel, and PLATO will expand our observational horizon, physically-consistent modelling provides the means to transition from categorical interpretations toward a comprehensive picture of the exoplanet continuum.
\end{abstract}


%%%%%%%%%%%%%%%%%%%%%%%%%%%%%%%%%%%%%%%%%%%%%%%%%%

\section{Introduction} 
\label{sec:intro}

All planets are physically-distinct objects with properties that arise from a collection of internal and external processes. However, we may attempt comparisons between seemingly-similar planets to gain zeroth-order insights into shared physical and chemical processes. Three intuitive and measurable variables for a given planet are its radius, mass, and orbital configuration. With these, we can place a planet within the wider population and constrain its bulk properties \citep{seager_mass_2007, Zeng2019}. For example, the sub-Neptune and super-Earth populations are distinguished by their radii ($R_p$) according to an orbital period-dependent sparsity in the exoplanet population around 1.7 Earth radii: the `radius valley' \citep{ho_shallower_2024, fulton_california-_2018, owen_evap_2017}.

Theoretical relationships between planet radii and masses have allowed inferences as to exoplanet composition and structure \citep{luque_density_2022, valencia_diversity_2025}. A planet's observed radius is a dependent quantity ultimately arising from an arrangement of core, mantle, and atmosphere. Each of these components have inhomogeneous material and thermal states which together set a planet's observable characteristics \citep{baumeister_fundamental_2025, unterborn_nominal_2023, Zeng2019}. Thus, connecting observations of a planet to its bulk properties, interior phase state and composition \citep{nixon_magma_2025, benneke_jwst_2024}, is strongly dependent on modelling; modelling approximations fundamentally shape how we interpret exoplanet observations. Two sets of potentially-limiting model assumptions relate to planetary \textit{composition} and to atmospheric \textit{structure}. 

\subsection{Composition and mass-radius}
Existing mass-radius libraries typically address three compositional scenarios: (1) airless planets without atmospheres, (2) rocky interiors enveloped by `primordial' \ch{H2}-\ch{He} atmospheres, and (3) volatile-rich structures composed of various \ch{H2O} phases \citep{adams_OceanPla_2008,lopez_understanding_2014,  dorn_structure_2015, Zeng2019, otegi_revisited_2020, turbet_revised_2020, huang_magrathea_2022}. The second case is based on analogy to giant planets with \ch{H2}-rich atmospheres with minor enrichment in metals \citep{owen_evap_2017}, while the third case is often justified by formation beyond the \ch{H2O} ice line \citep{Burn2024}.

However, the ice line is \textit{external} to the soot line, so planets forming \ch{H2O}-rich may be enriched in carbon-bearing refractories \citep{bergin_exoplanet_2023,li_soot_2025, boitard_iceline_2025}. Hydrogen-bearing volatiles also readily partition into planetary interiors, favouring the formation of carbon-rich atmospheres \citep{bower_retention_2022, nicholls_redox_2024}. Modelling of disk chemistry also suggests that some planets can incorporate substantial sulfur inventories which reflect their formation scenario \citep{kama_abundant_2019,  jorge_forming_2022, sossi_review_2025}. High-metallicity atmospheres are a common outcome of population synthesis modelling \citep{fortney_framework_2013,mordasini_formation_2015}, modulated by non-linear disk condensation sequences \citep{kimura_predicted_2022,spaargaren_proto_2025,zaveri_chemical_2026}. The true range of atmosphere diversity is often not incorporated into static-structure modelling.  

JWST observations provide empirical grounds for atmospheric compositions more diverse than those accessible to existing static structure approaches. The ultra-short period exoplanet 55-Cnc\,e ($M_p=7.99 M_\oplus$) may host a hydrogen-poor secondary \ch{CO2}-\ch{CO} atmosphere \citep{hu_55cnc_2024} sustained by outgassing from a deep dayside magma ocean \citep{Meier2023, Heng2023, bourrier_55cnc_2018}. On the basis of planetary evolutionary simulations, \citet{nicholls_escape_2025} showed that JWST measurements of super-Earth L\,98-59\,d ($M_p=1.64M_\oplus$) are consistent with its bulk composition comprising order~$\sim1\%$ sulfur via favourable sequestration of $\mathrm{S}^{2-}$ into the planet's deep interior \citep{gressier_hints_2024, cadieux_detailed_2025}. The super-Earth \PiMen\,c ($M_p=3.63M_\oplus$) has been argued to have a thick atmosphere comprised of $>50\%$ volatiles, justified by the detection of C\,II ions overflowing its Roche lobe \citep{Huang_pimenc_2018, Munoz_pimenc_2021, Hatzes_pimenc_2022}. On the other hand, the warm sub-Neptune TOI-421\,b ($M_p=6.7M_\oplus$) is thought to have an atmosphere of Solar metallicity: a potential example of the `primordial' \ch{H2}-\ch{He} end-member scenario \citep{carleo_toi421b_2020, davenport_toi421b_2025}. Taken together, these observations suggest that considering only the \ch{H2}--\ch{H2O} end-member scenarios will bias our interpretations of planet bulk properties, and their corresponding formation scenarios, on the population-level.  

\subsection{Atmospheric structure and mass-radius}
The direct impact of assumed atmospheric \textit{structure} on inferences of bulk planet properties also remains unexplored. While choices of interior structure and mantle make-up have been a concern of modelling accuracy in prior studies \citep[e.g.,][]{Schulze_TowardReli_2026}, the choice of atmospheric structure is important because it depends strongly upon temperature, pressure, and composition. Initial static-structure calculations \citep{seager_mass_2007} did not include the contribution of planetary atmospheres, but follow-up studies considered \ch{H2}-\ch{He} atmospheres and found that these envelopes inflate planet radii by up to $60\%$ \citep{adams_OceanPla_2008}. Several recent models do not account for atmospheric contributions to observed radii \citep[e.g.][]{unterborn_nominal_2023,plotnykov_Evidence_2025},  yet studies which \textit{do} consider atmospheres often adopt substantially-simplified structures. For example, widely-applied mass-radius calculations derived by \citet{Zeng2019} incorporate isothermal atmosphere structures. In reality, it is known that planetary atmospheres are not guaranteed to be polytropic, nor fully-convective, nor isothermal \citep{marley_review_2015, selsis_cool_2023, nicholls_convective_2025, peng_puffy_2024, cmiel_structure_2025}. Recently, while adopting pure-\ch{H2O} adiabatic atmospheres, \citet{Chakrabarty_TheRadius_2026} suggested that the small exoplanet census can be explained by a population of worlds comprising up to 46~wt\%~\ch{H2O} in bulk. However, without applying atmospheric models of suitable physical complexity in the structure calculations, there is a risk that such inferences alias other processes.

Another possible inaccuracy is how gravity is treated in atmospheres: Existing models adopt a range of approaches to approximate gravitational acceleration ($g$) as a function of height.  Gravitational self-attraction within massive sub-Neptune atmospheres decreases their radii sufficiently to enable atmosphere retention, so non-linear variation of $g$ as a function of atmosphere \textit{size} may be foundational to the exoplanet radius valley \citep{owen_evap_2017,owen_review_2019}. Departure from the $g\approx\mathrm{const}$  has also been shown to undermine the universality of a runaway greenhouse limit \citep{arnscheidt_Atmospher_2019}.  \citet{Molliere_petitRADTRA_2019} found that even a first-order $g\propto 1/r^2$ dependence yields substantial and observable differences in synthetic transmission spectra calculated by radiative transfer models. Despite this sensitivity, codes commonly utilised within spectroscopic retrieval frameworks (e.g., \citet{Molliere_petitRADTRA_2019,Kitzmann_heliosR_2020}) adopt a zeroth-order approximation of $g=\mathrm{const}$.  

Alongside a planet's mass and composition, its observable radius is also a function of its irradiation environment \citep{swift_massradius_2011,enoch_factors_2012}. A careful consideration of a planet's radiation environment may mitigate the substantial degeneracies in constraining its bulk composition: a given mass-radius pairing could be consistent with a wide range of compositions, but only a narrower range of scenarios may be physical when simultaneously considering its instellation \citep{ikoma_constraints_2006}. Rapid structure calculations remain valuable for  leading-order inferences of planet composition and surface environmental conditions from estimated bulk properties (e.g., for observation proposals), yet no tooling currently exists for realistically and efficiently relating observable parameters and irradiation environments to planet structure and environment. 

Parametrisation of a planet's stellar radiation exposure underpins the `habitable zone' (HZ) concept, which -- similarly to mass-radius relations -- provides a first-order framework for categorising exoplanets \citep{kasting_hz_1993, kopparapu_habitable_2013}. However, traditional habitable zone boundaries are  defined for Earth-sized planets independently of atmosphere mass fraction. Since massive hydrogen-dominated atmospheres efficiently blanket planetary interiors  and induce a strong greenhouse effect \citep{pierrehumbert_hydrogen_2011, nicholls_redox_2024}, this makes the traditional habitable zone boundaries inappropriate for comparison against sub-Neptunes \citep{innes_runaway_2023, luu_can_2024}. Unification of the habitable zone and mass-radius interpretive frameworks is methodologically feasible -- given sufficiently accurate atmosphere modelling -- and can, in principle, provide a robust connection between sub-Neptune climate states and observationally accessible parameters.

Previous studies have avoided on-the-fly evaluation of expensive forward models with pre-computed model grids \citep{husser_new_2013,fisher_retrieval_2018, goyal_library_2020, fisher_retrieval_2020}. While efficient for fitting spectroscopic measurements, these `grid-retrieval' approaches treat the atmosphere as being isolated from the deep interior. The existing \texttt{spright} code \citep{parviainen_spright_2024} deploys a Bayesian method to adaptively combine `mixtures' of compositional (hydrogen or water) and structural (rocky or icy) end members, representing an alternative framework to pre-defined mass-radius relations. However, \texttt{spright}'s authors note that integral atmosphere modelling simplifications within \texttt{spright} present substantial caveats when drawing conclusions from its probabilistic inferences \citep{parviainen_spright_2024}. The success of fast Bayesian internal-structure retrievals is also demonstrated by the \texttt{plaNETic} code \citep{egger_unveiling_2024}, although with modelling applicability restricted by their adopted planet models comprising tripartite water (steam), silicate, and iron mixtures \citep{haldemann_aqua_2020}.

\subsection{Paper outline}
Here, we address the arising tension between simplified planetary structure modelling, known correlations between the relevant physics, and the diverse atmospheric compositions evidenced by recent observations. We use radiative-convective atmosphere models to assess the sensitivity of observable exoplanet structure to a diversity of atmospheric compositions and energy transport processes. In doing so, we present a comprehensive library of planet structures which simultaneously unifies mass-radius structure relations with the habitable zone framework. With this library of planet structures, we then develop a rapid Python-based retrieval tool through a Bayesian-inference approach. We unify prior `grid-retrieval' methodologies \citep{fisher_retrieval_2020, barstow_outstanding_2020} with simulations of whole-planet structure, thereby incorporating structural constraints on photospheric observables. Coupling radiative-convective-chemical equilibrium climate modelling with an interior model ensures that comparisons with observations can remain physically self-consistent with total planetary mass, radius, energy budget, and internal structure.

The structure of this paper is as follows:
\begin{enumerate}
    \item Methods sections~\ref{sec:methods_interior},~\ref{sec:methods_compose},~and~\ref{sec:methods_planet} describe our statically coupled interior-atmosphere model, founded upon previously-validated codes, which connects atmosphere climate calculations to planetary interior structures. We also outline  the calculated model outputs variables, derived from its input parameters.

    \item Methods section~\ref{sec:methods_params} develops a library of planetary structures using radiative-convective atmosphere modelling across a wide parameter space. Methods section~\ref{sec:methods_infer} outlines the static model's incorporation into a new open-source Bayesian-inference tool.

    \item Section~\ref{sec:results_atmosphere} considers a case study to probe how atmospheric temperature structure simplifications can, in principle, influence planet radii.
    
    \item Section~\ref{sec:results_curves} presents our library of models and identifies physics-driven systematics. Sections~\ref{sec:results_gravity}~and~\ref{sec:results_convect} empirically assess variations in gravitational attraction and convective strength, to probe how these physical processes may systematically deviate from simplified scenarios. 
    
    \item Section \ref{sec:results_hz} considers the broader radiation-energy \textit{environment} in which planets are ultimately located to empirically evaluate sub-Neptune habitability in the context of physical systematics, and unify planet structure relations with the habitable zone concept. 

    \item Section~\ref{sec:results_ia} demonstrates rapid Bayesian retrievals founded upon our library of models, adopting the cases of TOI-421\,b and \PiMen\,c as testable examples.

    \item Discussion section~\ref{sec:discuss} outlines the implications of our results for theorists, highlights critical sensitivities which must be considered in light of upcoming telescopes for observers, and describes a reliable pathway for future exoplanet characterisation.
    
\end{enumerate}

\section{Methods} 
\label{sec:methods}

\subsection{Interior structure}
\label{sec:methods_interior}
This work focuses on atmospheric structure. However, our models require a planetary interior solution. We derive an analytical fit for interior radius $R_s(M_s,f_{ci})$ (where $f_{ci}=M_c/M_s$ is the metallic-core mass fraction and $M_c$ is the core mass) to facilitate our subsequent methodology.

Solutions of interior radius $R_s$ are obtained for interior masses $M_s$ between 0.25~and~$10.25\text{ M}_\oplus$ and core interior-mass fractions $f_{ci}$ between 0.10~and~0.75. The internal structures of density, gravity, and radius are solved iteratively by integrating a system of three coupled differential equations until a solution which conserves mass is attained:
\begin{align}
    \dd M / \dd r &= 4\pi r^2 \rho \\
    \dd g / \dd r &= 4\pi G \rho - 2g/r \\
    \dd p / \dd r &= -\rho g
\end{align}
These three equations describe conservation of mass, gravitational attraction, and hydrostatic equilibrium, respectively.

We consider an interior model comprising a fully-differentiated iron core and silicate mantle and use the equation of state described by \citet{seager_mass_2007}, which is valid for planet masses up to $50M_\oplus$. The well-established \citet{seager_mass_2007} formalism treats the temperature-dependence of mantle material properties as a thermal pressure contribution, so hydrostatic pressure predominantly controls the density of mantle material. As such, we approximate the interior structure as being independent of temperature and completely solidified \citep{seager_mass_2007, bower_linking_2019}. These simplifications are justified because variations in mantle melt fraction and mineral assemblage at high pressures $>3\mathrm{\,GPa}$ are known to have only minor effects on the density of mantle material \citep{seager_mass_2007, unterborn_nominal_2023, boley_fizzy_2023}. Typical deep-mantle adiabatic geotherms cross both the liquidus and solidus melting curves of \ch{MgSiO3} rapidly with increasing pressure, so the deep mantles of super-Earth and sub-Neptune planets will generally be solidified at their high-pressure conditions \citep{Andrault_Solidusand_2011, wolf_equation_2018}. Similarly, the storage of volatiles, \ch{Fe}, and \ch{FeO} within the mantle are known to have only a negligible impact on planetary interior structure  \citep{unterborn_nominal_2023}. 

\begin{figure}
    \centering
    \includegraphics[width=0.92\linewidth]{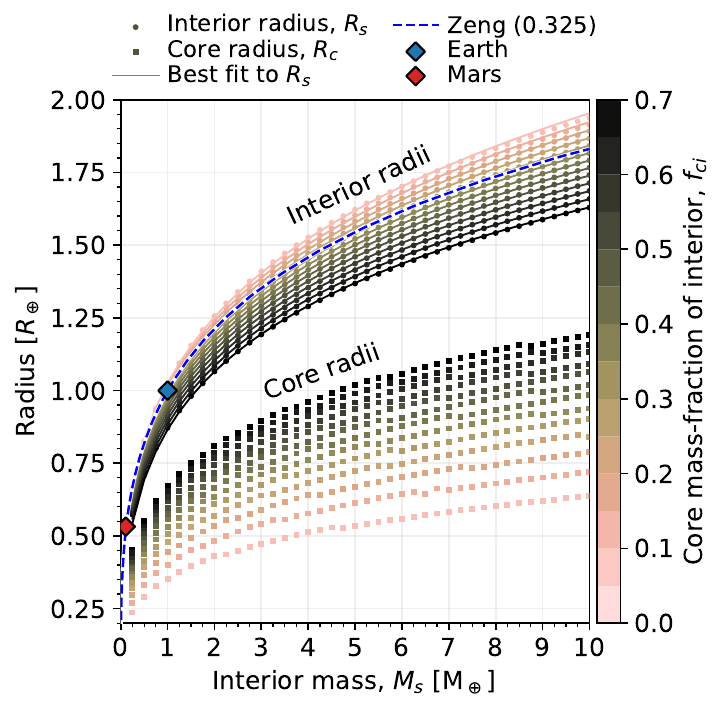}%
    \caption{Interior radii (circles) and core radii (squares), as a function of total planet mass and core interior-mass fractions (colour). Analytical fits $R_s(M_s, f_{ci})$ to these models are shown by solid lines. The  mass-radius relations for an Earth-like core mass fraction $f_{ci}=0.325$ calculated by \citet{Zeng2019} is shown by the dashed blue line. Solar System planets included for reference (diamonds).}
    \label{fig:interior}
\end{figure}

The interior radius, $R_s$, is calculated as a function of interior mass $M_s-$ and core interior-mass fraction $f_{ci}$ (Figure \ref{fig:interior}). 

An analytical function is fitted to these $R_s$ samples using seven constants,
\begin{equation}
    \label{eq:interior_only}
    R_s(M_s, f_c) = \alpha_0 M_s^{\alpha_1} + \beta_0 f_{ci}^{\beta_1} + \gamma_0(M_s\cdot f_{ci})^{\gamma_1} + \delta ,
\end{equation}
where $R_s$ and $M_s$ are in units of $R_\oplus$ and $M_\oplus$, and $f_{ci}$ is the interior-mass fraction of the metallic core. The best-fitting solution, rounded to 5 decimal places, is: $\alpha_0 = +1.20345$, $\alpha_1 = +0.26380$, $\beta_0 = -0.21157$, $\beta_1 = +1.99228$, $\gamma_0 = -0.10285$, $\gamma_1 = +0.59099$, $\delta = -0.15051$. For these constants, the fit has a coefficient of determination $R^2=0.99985$ and mean absolute percentage error $\mathrm{MAPE}=0.21870\%$. This fit is shown by solid lines in Figure \ref{fig:interior}. 

An atmosphere of mass $M_a$ and mass fraction $f_a=M_a/M_p$ can be algebraically incorporated into Equation \ref{eq:interior_only}. Given that the total mass $M_p$ can be divided between these three components (core, mantle, and atmosphere), $M_p = M_c + M_m + M_a$, we can write the core interior-mass fraction $f_{ci}$ in terms of the core planet-mass fraction $f_c$ as $f_{ci}= f_c / (1-f_a)$. Doing so allows us to incorporate the atmosphere mass $M_a=f_aM_p$ into our interior structure fit, on the assumption that the mantle and core are negligibly compressed by an overlying atmosphere \citep{bower_linking_2019, krissansen-totton_waterworlds_2021}. 

We can retain the same $\alpha\beta\gamma\delta$ coefficients while incorporating the atmosphere mass into this fit, yielding
\begin{equation}
    \label{eq:interior}
    R_s = \alpha_0 \Big[(1-f_a)M_p \Big]^{\alpha_1} + \beta_0 \Big[\frac{f_c}{1-f_a} \Big]^{\beta_1} + \gamma_0 \Big[f_c M_p \Big]^{\gamma_1} + \delta.
\end{equation}
It is possible to calculate a planet's interior radius $R_s$ directly from its total mass $M_p$ and component mass fractions ($f_c$ and $f_a$) using  Equation \ref{eq:interior}.

From the spread of coloured lines in Figure \ref{fig:interior} across $R_s$-space (y-axis), it is apparent that wide variations in metallic core interior-mass fraction (line colour) can generate differences in planet radius of no more than $0.35 R_\oplus$, for a large $10 M_\oplus$ planet. So far, these variations in interior radius neglect any atmospheric radial contribution.

\subsection{Atmosphere composition}
\label{sec:methods_compose}
We calculate whole-planet radial structures by placing atmosphere models atop the interior models described by Equation \ref{eq:interior}. To do this, we must first parametrise the atmospheric composition. The total surface pressure -- which is agnostic of any particular atmospheric composition -- is directly related to atmospheric mass fraction $f_a$ and total planet mass $M_p$ via
\begin{equation}
    \label{eq:psurf}
    p_s = \frac{g_sf_aM_p}{4\pi R_s^2} ,
\end{equation}
where the gravity $g_s$ at the surface is given by
\begin{equation}
    \label{eq:gravity}
    g_s = G (1-f_a)M_p / R_s^2,
\end{equation}
where $G$ is Newton's gravitational constant.

Numerous physical and chemical processes control atmospheric composition. Magma ocean degassing is likely a primary controller, but its particular behaviour remains poorly understood on both theoretical and empirical grounds \citep{elkins_linked_2008,schaefer_review_2018,sossi_redox_2020}. Regardless of volatile delivery and partitioning mechanisms, atmospheric speciation is subject to chemical processing \citep{madhusudhan_exoplanetary_2016, nicholls_temperaturechemistry_2023, werlen_atmospheric_2025}.
Here, the gaseous CHONS composition is calculated at thermochemical equilibrium, assuming that the atmosphere is \textit{elementally} well-mixed throughout the column. The CONS abundances, as mass ratios relative to hydrogen, are derived from two input parameters: the total metallicity $Z_a$ and the C/O~ratio. The total metallicity is defined as
\begin{equation}
    Z_a = \frac{m_C}{m_H} + \frac{m_O}{m_H} + \frac{m_N}{m_H} + \frac{m_S}{m_H},
    \label{eq:Za}
\end{equation}
where $m_e$ are the total masses (kg) of each element's reservoir in the atmosphere. Rearranging Equation \ref{eq:Za} for the elemental mass mixing ratios of C and O relative to H yields
\begin{align}
    y_C &= \frac{m_C}{m_H} = \frac{\mathrm{C/O}}{1+\mathrm{C/O}}\Big( Z_a - y_S-y_N\Big), \nonumber \\
    y_O &= \frac{m_O}{m_H} = \frac{y_C}{\mathrm{C/O}}.
    \label{eq:yCO}
\end{align}
These mass ratios are converted to the logarithmic molar abundances
\begin{equation}
    x_e=\log_{10}(y_e\mu_H/\mu_e)+12
    \label{eq:xe}
\end{equation}
for our thermochemical equilibrium modelling (Section~\ref{sec:methods_planet}). The solar photosphere is $Z_\odot=0.0314$, corresponding to $\log_{10} Z_a=-1.9$ in the nomenclature of our model parameters.

Equations~\ref{eq:Za}~and~\ref{eq:yCO} enable the exploration of total atmospheric metallicity and C/O ratio as independent parameter axes which vary the atmospheric abundances of CHO elements.  However, other volatile elements are expected to be present within planetary atmospheres \citep{sossi_review_2025, krijt_chemical_2023} and spectroscopic JWST observations have inferred the presence of sulfur-species within sub-Neptunes and hot-Jupiters \citep{tsai_photochemically_2023,fu_hydrogen_2024,gressier_hints_2024,  belloarufe_evidence_2025,  nicholls_escape_2025, dai_photochemic_2026}.  Entirely neglecting S and N from our chemical calculations would be unrepresentative of real atmospheres, although we  build upon prior planet-structure simulations by considering how $Z_a$, C/O, temperature, and pressure physically determine the CHO speciation.  We therefore strive to incorporate S and N species into our  modelling -- while limiting our parameter space and scope -- by considering conservative lower bounds on atmospheric S and N abundances. The Earth is depleted in volatile elements compared to the Sun's photosphere, relative to rock-forming elements \citep{wang_volatility_2019, sossi_review_2025}, but the metallicities of Earth's volatile elements (e.g., C/H, S/H) are \textit{enhanced} compared to their equivalent in the Solar photosphere ratios \citep{hirschmann_constraints_2016}. Solar-like metallicities can be adopted as conservative lower-estimates on the S and N abundances expected within exoplanetary atmospheres. So, here we adopt solar photosphere values of $x_S=7.12$ and $x_N=7.83$ for our thermochemical calculations \citep{asplund_chemical_2009}. These are held constant self-consistently with the parameters $Z_a$ and C/O, via Equation \ref{eq:yCO}.

\subsection{Atmosphere structure}
\label{sec:methods_planet}

The atmospheric temperature and height structure is solved self-consistently at radiative-convective-chemical equilibrium using AGNI: an established 1D hydrostatic atmosphere model, which incorporates real-gas equations of state, spectral radiative transfer, mixing-length convection, and gas-phase thermochemistry \citep{nicholls_agni_2025, nicholls_convective_2025}. 

Using AGNI, we self-consistently solve for the atmospheric temperature, pressure, radius, gravity, and composition profiles using a damped Newton-Raphson optimisation scheme \citep{press_numerical_2007}.  Damping is performed using a backtracking linesearch method \citep{press_numerical_2007}. The iterative Newton-Raphson method is initialised with a log-linear $T(p)$ profile, and then iterates climate-chemistry profiles, using forward-model evaluations of energy fluxes which construct a Jacobian matrix using a second-order central difference scheme. We thereby iteratively solve for the equilibrium atmosphere state, minimising the 4-norm of energy flux losses throughout the atmosphere column, conserving energy through each layer up to a small tolerance that represents our `strong' convergence criterion, \citep{nicholls_agni_2025}.  Our method for solving for energy-conservation in the atmosphere represents radiative-convective-chemical-hydrostatic equilibrium \citep{nicholls_convective_2025, peng_puffy_2024}. The atmospheres are not explicitly time-stepped; as in previous studies, we assume that they attain chemical and thermodynamic equilibrium on timescales shorter than their present-day age \citep{schaefer_review_2018, turbet_revised_2020,aguichine_ocean_2021, krissansen_erosion_2024}.

In AGNI, water is treated with the AQUA equation of state \citep{haldemann_aqua_2020}, hydrogen with the \citet{chabrier_eos_2019} equation of state, other volatiles with the Van~der~Waals EOS where coefficients are available \citep{coker_thermo_2007} and with the ideal gas equation of state otherwise \citep{nicholls_escape_2025}. Gas densities are combined using Amagat's additive volume law, which assumes that each gas can exhibit non-ideal behaviour but they are combined `ideally' \citep{magyar_eos_2014, magyar_ethane_2015, dorman_eos_1991}.  These equations of state tabulations are applicable to pressures up to $10^6$\,bar and have been previously applied to sub-Neptunes and giant planets \citep{saumon_eos_1995, tremblin_thermo-compositional_2019, baraffe_new_2015}. We do not explicitly model the effects of immiscibility between supercritical phases, which is suggested to cause de-mixing and compositional stratification in sub-Neptune interiors \citep{rogers_redif_2025, gupta_the_2025}, but explore this possibility for the case of TOI-421\,b (Section~\ref{sec:methods_infer}).

The effectively-observable photospheric radius of the planet $R_p$ is obtained from our atmospheric structures by extracting the calculated hydrostatic radius corresponding to the \SI{20}{\milli\bar} pressure level \citep{lopez_understanding_2014, Zeng2019, fortney_planetary_2007}. We note that the photosphere region probed by transmission spectroscopy is known to also depend on atmospheric composition, temperature structure, and the wavelength being observed \citep{aguichine_ocean_2021, valencia_diversity_2025}. In general, the photospheric pressure is expected to vary from 1\,mbar to 100s\,mbar depending on these factors \citep{fortney_transit_2005, dotson_exoplanets_2011}.  Care must also be taken when comparing these representative whole-planet climate structure calculations against the limb profiles probed by transit measurements \citep{hubbard_transit_2001, burrows_transit_2003, grant_trans_2023}. To address these issues, we explore the relationship between photosphere location and atmosphere structure through an additional analysis in Appendix~\ref{app:transspec}. These calculations demonstrate that defining the photosphere at  20\,mbar is appropriate for comparison of the model with measurements from JWST and PLATO \citep{alatalo_jwst_2016, rauer_plato_2025}.

The atmosphere column is defined on a pressure grid with $N=40$ layers between a fixed 1\,mbar top-of-atmosphere pressure and an adaptive surface pressure, which is calculated from Equation~\ref{eq:psurf} in each modelled scenario. The atmosphere layers are arranged logarithmically in pressure-space, except at the top- and bottom-boundaries, where the layers are kept numerically small to obviate numerical boundary-layer effects. Each iteration of the Newton-Raphson scheme determines a new estimate for the temperature profile, from the Jacobian matrix, which is then used to calculate gas-phase composition profiles (with FastChem as described below), thermodynamic properties (using our equation of state formalism), and hydrostatic structure. The hydrostatic pressure-height-gravity integration is performed from the surface  upwards by simultaneously integrating the gravity equation,
\begin{equation}
    g(r)=G\frac{M(r)}{r^2},
    \label{eq:grav_enclosed}
\end{equation}
the hydrostatic equation,
\begin{equation}
    \dd p = -\rho(p,T)g(r) \dd r,
    \label{eq:hydro_pres}
\end{equation}
and the mass continuity equation,
\begin{equation}
    \dd M(r) = \frac{1}{g(r)}\dd p(r),
    \label{eq:hydro_mass}
\end{equation}
using a fourth-order Runge-Kutta method on 2000 layers. The surface boundary conditions are $g(R_s)=g_s$, $p(R_s)=p_s$, and  $M(R_s)=M_s$ (Section~\ref{sec:methods_interior}). Densities $\rho$ and thermodynamic properties are interpolated from the 40 pressure layers to the 2000 height layers, using a PCHIP scheme, at each solver iteration \citep{fritsch_pchip_1984}. The height and gravity profiles obtained from hydrostatic integration of Equations~\ref{eq:grav_enclosed}--\ref{eq:hydro_mass} are interpolated back onto the 40 layer pressure grid for our climate and energy-flux calculations. We account for the gravitational contribution of \textit{all} mass enclosed within each layer of the planet's structure, including attraction of the upper atmosphere towards the deeper atmosphere layers in addition to the interior.

The thermochemical equilibrium speciation of CHNOS elements into molecules is calculated at each atmosphere pressure-layer using the FastChem3 code, according to the local temperature and pressure conditions \citep{ stock_fastchem_2018,stock_faschem_2022}. The atmospheric \textit{elemental} composition -- set by $Z_a$ and C/O -- is constant throughout the atmosphere column in each case, and provided to FastChem as CNOS atomic number densities normalised relative to the hydrogen atom number density (Equation \ref{eq:xe}). The gas compositions calculated from FastChem are used to calculate the thermodynamic properties and densities at each atmosphere layer \citep{JANAF, coker_thermo_2007}.

Each solver iteration evaluates energy fluxes from the atmospheric temperature, composition, and height profiles. Spectroscopic radiative fluxes are  modelled with SOCRATES \citep{manners_socrates_2024, edwards_studies_1996} under the two-stream approximation using a correlated-$k$ method for parametrising gas opacities. Line opacities for \ch{H2O}, \ch{H2}, \ch{CO2}, \ch{CO}, \ch{CH4}, and \ch{N2} are obtained from the DACE database, which are largely drawn from ExoMol, using a 25\,cm$^{-1}$ line wing cut-off \citep{grimm_database_2021,tennyson_exomol_2018}. Continuum opacities are obtained from the HITRAN CIA database \citep{karman_hitran_2019}, except \ch{H2O} uses the semi-empirical MT\_CKD version 3.2 continuum parameterisation \citep{mlawer_mtckd_2012,mlawer_mtckd_2023}. All other gases formed by thermochemistry are treated with zero opacity. We include Rayleigh scattering, but neglect aerosols and clouds.  The top boundary condition on our radiative transfer calculations is set by the down-welling stellar radiation, treated as a blackbody function of stellar effective temperature $T_\text{eff}$ and bolometric instellation $F_\text{bol}$.  

Atmospheric convection is parametrised using Schwarzschild mixing-length theory, which assumes that unstably-buoyant parcels of ascending gas  diffuse energy over a characteristic mixing length $\lambda$ \citep{ vitense_mlt_1953,  robinson_temperature_2014, joyce_mlt_2023}. The mixing-length asymptotically tends to the pressure-scale height aloft and to zero at the surface \citep{hogstrom_karman_1988}. Importantly, our mixing-length formalism estimates the characteristic convective velocity $w$ from the Brunt-V\"as\"all\"a frequency of buoyant parcel oscillations   \citep{pierrehumbert_book_2010, windsor_radiative-convective_2022}, and can thereby derive a  vertical eddy diffusion coefficient,
\begin{equation}
K_{zz} = w\lambda.
\end{equation}
Our estimates of $K_{zz}$ enable an assessment of dynamical mixing processes within these atmospheres, in Section~\ref{sec:results_convect}.

The upward convective energy flux is combined with the net radiative flux to determine the total transport of energy through each layer of the model. This formalism naturally permits the formation of convectively-stable layers -- where energy is transported entirely by radiation -- which are recognised to strongly modulate surface temperature conditions and atmospheric structure, at a given instellation \citep{selsis_cool_2023,  peng_puffy_2024, cmiel_structure_2025}.

There are two options for setting our atmosphere energy bottom-boundary condition: (a) enforce a surface temperature $T_s$, or (b) enforce a net heat flux $F_\text{net}$. The $T_s$ case (a) could be interpreted as a planet which is initially hot after formation, far from radiative equilibrium, with a correspondingly large $F_\text{net}$ \citep{elkins_linked_2008, nicholls_redox_2024}. The $F_\text{net}$ boundary condition (b) is identical to imposing a constant internal temperature ($F_\text{net}=\sigma T_\text{int}^4$) and could be interpreted as a steady-state with internal heat production balanced by cooling to space \citep{lopez_understanding_2014, hamano_emergence_2013, nicholls_redox_2024, van_onset_2025}.  We adopt boundary condition (b) with $F_\text{net}=\SI{0.1}{\WPMS}$ in our main grid  models, representative of Earth's present geothermal heat flux, and a conservative lower-bound on the internal heat production expected within super-Earth and sub-Neptune exoplanets \citep{lodders_planetary_1998, stevenson_interiors_1982, welbanks_Ahighin_2024}.

\subsection{Library of planetary structure models}
\label{sec:methods_params}
We produce a large library of planetary structure models which span a parameter space encompassing a range of structure configurations, atmospheric compositions, and irradiation environments (Table \ref{tab:params}). The foremost grid axis is the total planet mass $M_p$, which can be derived from radial velocities using Doppler spectroscopy or through transit timing variations \citep{dotson_exoplanets_2011, perryman_exoplanet_2018}. We explore masses $M_p$ spanning scenarios from the terrestrial to the sub-Neptunian regimes.

A planet's total mass can be considered to arise from the sum of three components: a core, a mantle, and an envelope (atmosphere). Although we focus primarily on atmospheric structure in this present work, we necessarily also consider a small set of core mass fractions $f_c$ from $0.2$ to $0.7$. This range encompasses small $f_c$ values inferred for rocky exoplanets \citep{Zeng2019} and for Mars ($f_c\sim0.23$; \cite{khan_mars_2018}), as well as Earth's value ($f_c\sim0.325$; \cite{lodders_planetary_1998}), and extends to the large core mass fraction estimated for Mercury ($\sim0.7$; \cite{hauck_curious_2013}). 

Volatiles are expected to partition between planetary atmospheres and interiors. However, dissolved volatiles are expected to have only a minor effect on planetary internal density structure, and thus a planet's interior radius at a given mass. We do not perform volatile partitioning calculations here because there are substantial uncertainties and known-unknowns in volatile partitioning behaviour across the full planet mass range we investigate \citep{gaillard_diverse_2021, sossi_solubility_2023}. Furthermore, the capacity for planetary interiors to store volatiles may not be limited by solubility, as hydrogen and silicates may be miscible within sub-Neptune mass planets \citep{young_diff_2025, rogers_redif_2025}. We implicitly incorporate these effects into our atmospheric parameters ($f_a$, $Z_a$, C/O), which are agnostic of any specific formation pathway. For a given total mass $M_p$ and planet radius $R_p$, atmosphere masses $f_a$ considered here may be interpreted as lower-bounds on the planet's \textit{total} volatile inventory \citep{DornLichtenberg2021}. 

Our grid incorporates atmospheric mass fractions ranging logarithmically from a minimum value of  $f_a=10^{-3.25}$, up to a 25\,wt\% scenario in which the atmosphere comprises a substantial fraction of the whole-planet mass. Our grid is regularly sampled, so the sum of atmosphere and metallic-core masses must be less than the total planet mass. That is, $f_a$ must always satisfy $f_c+f_a<1$, with the remaining mass corresponding to the planet's silicate mantle layers. 

The atmosphere mass fraction $f_a$ only determines the \textit{total} surface pressure through Equation~\ref{eq:psurf}, but remains agnostic to the atmospheric speciation. Alongside $f_a$, we also consider two atmospheric composition axes to solve for the gas mixing ratios at thermochemical equilibrium. The first variable determining the atmospheric elemental composition is its total metallicity $\log_{10} Z_a$ (Equation \ref{eq:Za}), varied from -2 to +1, which controls the mass ratio of hydrogen to `metals'. The second variable determining the atmospheric elemental composition is the $\log_{10} \mathrm{C/O}$~ratio (Equation \ref{eq:yCO}), varied from -3 to 0. These metallicity variables are provided to FastChem to solve for the gas-phase thermochemical composition, using Equation~\ref{eq:xe}. Solar metallicity corresponds to $\log_{10} Z_a=-1.87$ \citep{asplund_chemical_2009}. The domain of atmospheric compositions that a real planet could draw from, at chemical equilibrium, is thus encompassed by our wide parameter space, generalised across all possible scenarios by which these atmospheres may be generated (e.g., by magma ocean degassing; \citet{werlen_atmospheric_2025}).

The observational techniques used to determine a planet's mass $M_p$ also provide a solution for its orbital semimajor axis $a_p$. The planet's bolometric instellation flux $F_\mathrm{ bol}$ is indirectly described by its orbital semimajor axis $a_p$ and the integrated bolometric luminosity $L_\mathrm{bol}$ of its host star,
\begin{equation}
    F_\mathrm{bol} = L_\mathrm{bol}/(4\pi a_p^2).
\end{equation}
For our radiative transfer calculations, we treat the spectroscopic stellar emission as a blackbody -- evaluated using the Planck function -- with the star's photospheric emission represented by an effective temperature:
\begin{equation}
    L_\mathrm{bol}=\sigma T_\mathrm{eff}^4.
\end{equation}
Together, parameter axes $T_\text{eff}$ and $F_\mathrm{bol}$ define the planetary irradiation environment and define a planet's location relative to its host star's habitable zone. We quantify the bolometric instellation flux in units of `solirads' \citep{Mamajek_TheSolirad_2025}, where \SI{1}{\soli} equals Earth's average instellation flux of $S_\oplus = \SI{1361}{\WPMS}$ \citep{kopp_solar_2011}. We consider a range of irradiation scenarios with bolometric fluxes varied from $F_\mathrm{bol}=$1\,So~to~1000\,So and stellar temperatures varied from $T_\mathrm{eff}=2500$\,K~to~6500\,K, informed by the small-exoplanet census (Appendix~\ref{app:exoplanets}).  For a planet orbiting a Sun-like star ($T_\mathrm{eff}=5772$\,K; \cite{gueymard_suns_2004}), the limits of our irradiation parameter space probe circular orbital separations from $a=0.032$\,AU~to~$1.000$\,AU and orbital periods from $P=365.22$\,days~to~$2.05$\,days. For a planet orbiting a TRAPPIST-1-like star ($T_\mathrm{eff}=2566$; \cite{agol_refining_2021}), the parameter space probes separations from $a=0.001$\,AU~to~$0.024$\,AU and orbital periods from $P=4.41$\,days~to~$0.03$\,days.  We thereby account for the role of stellar radiation in setting atmospheric structure (which is simultaneously sensitive to atmospheric opacity and composition) while providing a connection between the habitable zone concept and measurable planetary properties which arise from its structure. 

\begin{table}
\caption{Parameters varied across our grid of models, with corresponding ranges. Some are varied with linear sampling, some logarithmically, and others use specific values. A total of 504,000 simulated planets.}
\centering
\label{tab:params}
\begin{tabular}{lll}
    \hline \hline
    Parameter                   & Range   & Sampling\\ 
    \hline
    Total planet mass, $M_p$              & 1.0  -- 10.0$M_\oplus$    & 15, log \\
    Core mass fraction, $f_c$       & 0.2  -- 0.7               & 5, lin  \\
    Atmos. mass fraction, $f_a$       & $10^{-3.25}$ -- 0.25      & 8, log  \\
    Atmos. metallicity, $Z_a$         & $10^{-2}$ -- $10^{+1}$    & 7, log  \\
    Atmos. C/O~mass ratio           & $10^{-3}$ -- $10^{0}$     & 4, log  \\
    Instellation flux, $F_\text{bol}$    & 1 -- 1000\,So             & 5, log  \\
    Star temperature, $T_\text{eff}$& 2500  -- 6500 K           & 6, lin  \\
\end{tabular}%
\end{table}

For all models, we record the variables outlined in Table \ref{tab:output}. These include a measure of atmospheric convective vigour (maximum eddy diffusion coefficient across each profile, $K_{zz}$), the location of any convective zone, and the radius at which the atmosphere becomes weakly bound by gravity $R_b$ (where gravitational attraction $g$ is less than \SI{e-4}{\mpss}). The mean molecular weight $\mu_p$, temperature $T_p$, and gravity $g_p$ at the photosphere permit calculation of the scale height at that level $H_p=RT_p/(\mu_pg_p)$, which determines the relative size of absorption features probed by transmission spectroscopy \citep{barstow_outstanding_2020}. Only temperature, pressure, and radius are retained with resolved vertical profiles, due to the large number of grid points produced. However, these three recorded 1D-quantities allow complete reproduction of all other calculated quantities. 

\begin{table}
\caption{Output variables recorded at each grid point. The final four rows (italicised) describe vector quantities.}
\centering
\label{tab:output}
\begin{tabular}{llll}
    \hline \hline
    Variable     & Units        & Physical meaning & Scaling  \\ 
    \hline
    $p_s$        & bar          & Total pressure at surface & log \\
    $T_s$        & K            & Temperature at surface & lin\\
    $R_s$        & $R_\oplus$   & Interior (mantle) radius & lin \\
    $\mu_s$      & \SI{}{\amu}  & Mean molecular weight at surface & lin \\
    $g_s$        & \SI{}{\mpss} & Gravity at surface & lin\\
      
    $T_p$        & K            & Temperature at photosphere & lin\\
    $R_p$        & $R_\oplus$   & Radius of photosphere & lin\\
    $\mu_p$      & \SI{}{\amu}  & Mean molecular weight at photosphere & lin \\
    $g_p$        & \SI{}{\mpss} & Gravity at photosphere & log\\

    $R_b$        & $R_\oplus$   & Radius of weak binding & log\\
      
    $K_{zz}$     & \SI{}{\cmsps}& Maximum $K_{zz}$ across profile  & log\\
    $p_c^t$      & bar          & Top of convective zone & log\\
    $p_c^b$      & bar          & Bottom of convective zone & log\\

    $F_\text{loss,max}$      & \SI{}{\WPMS}     & Max flux difference in atmosphere & lin\\
    $F_\text{loss,med}$      & \SI{}{\WPMS}     & Median flux losses in atmosphere & lin\\
    $F_\text{TOA}$      & \SI{}{\WPMS}          & Net flux at top of atmosphere & lin\\
    $F_\text{BOA}$      & \SI{}{\WPMS}          & Net flux at bottom of atmosphere & lin \\

    $\chi_s^g$   & 1           & \textit{Surface mixing ratios of gases} & both\\
    $F_b$        & \SI{}{\WPMS} & \textit{Emission fluxes in bands} & log\\
    $T(p)$       & K            & \textit{Temperature-pressure profiles} & lin\\
    $r(p)$       & \SI{}{\mpss} & \textit{Radius-pressure profiles} & lin\\
\end{tabular}%
\end{table}

Not all grid points simultaneously satisfy our criteria for radiative-convective energy balance and hydrostatic equilibrium. This outcome is a natural result of our regularly sampled grid, which necessarily probes scenarios in which massive low-metallicity atmospheres enveloping low-mass interiors are subject to extreme irradiation environments. Regular grid sampling is necessary for the methodology described in the following section. Overall, 425503 cases (84.4\%) strongly respect our primary `strong' convergence criterion, which requires that energy fluxes are conserved across every atmosphere layer. However, only 99.8\% of all visited grid points have median atmosphere flux losses less than \SI{1}{\WPMS}, which we take as an alternative `weak' convergence criterion. We record all scenarios and recognise these criteria appropriately to the analyses in each section of this paper.

\subsection{Bayesian inference tool}
\label{sec:methods_infer}

Direct comparisons of planet mass-radius against pre-defined  isolines in mass-radius space do not readily incorporate uncertainties on the measurements and marginalise over the possibly consistent physical scenarios \citep{nixon_methods_2024, nixon_new_2024, parviainen_spright_2024}. For example, both atmospheric metallicity $Z_a$ and mass fraction $f_a$ contribute to atmospheric extent, and thereby set the observed radius $R_p$ \citep{barstow_outstanding_2020}. To self-consistently account for the multiple parameters which determine planetary structure and observables -- as well as uncertainties on these measurements -- we apply a Bayesian inference method to derive posterior distributions on our parameters \citep{madhusudhan_bayes_2009}. 

We adopt a numerical forward model which evaluates the observable variables $\mathcal{M}$ of our grid (Table~\ref{tab:output}) as a function of the grid's parameters $\theta$ (Table~\ref{tab:params}). Firstly, the grid parameters and observables are non-dimensionalised and logarithmically scaled, as appropriate. An instance of SciPy's regular grid interpolator is then created for each observable variable, as a function of our parameters \citep{virtanen_scipy_2020}. These interpolators provide a sufficiently smooth function by which a retrieval tool can infer posterior distributions on Table~\ref{tab:params}'s parameters $\theta$. Interpolation is substantially more performant than computing bespoke climate calculations at each forward-model evaluation step. This approach still allows 1D radiative-convective-chemical planet structure models to be integrated into fast retrievals, so observations of upper-atmosphere compositions and bulk planet properties (parameters in Table~\ref{tab:params}) can provide information about the conditions deep within planetary interiors and throughout their atmospheres.

The posterior probability distribution $P(\theta \vert \mathcal{O})$ for the parameters vector  $\theta$, given the `true' observations $\mathcal{O}$, is sampled using the affine-invariant Markov Chain Monte Carlo (MCMC) ensemble sampler `emcee' \citep{foreman_emcee_2013,goodman_ensemble_2010}. The sampler within emcee assembles multiple walkers to explore the parameter space efficiently. At each walker step, the MCMC algorithm evaluates our forward model -- the interpolator mapping parameters $\theta$ to modelled observables $\mathcal{M}$ -- and compares these observables $\mathcal{M}$ to the `true' observations $\mathcal{O}$. We adopt a uniform prior $P(\theta)$, where the probability is constant within the bounds of the model grid (Table~\ref{tab:params}) and zero elsewhere. The log-likelihood function by which emcee samples the posterior space incorporates asymmetric Gaussian uncertainties on $\mathcal{O}$  \citep{Davey_Investigati_2025}. 

The walker computations are distributed across multiple CPU cores in parallel. For a standard retrieval, we employ 9000 steps per walker with a 2\% burn-in fraction. These parameters are adjustable by the user, based on the complexity of the specific planet's posterior. To reduce autocorrelation in the final chain, we apply a thinning factor before generating the final posterior distributions. Each variable's autocorrelation time  is assessed to ensure convergence upon a solution \citep{goodman_ensemble_2010}.

Our grid and this retrieval framework are both accessible through a Python package termed `InferAGNI', which we make available via GitHub and PyPI (Section~\ref{sec:dataavail}). After installing this package with \texttt{pip}, it can be readily utilised with pre-provided Jupyter Notebook workflows, directly through a command line interface \texttt{inferagni}, or by importing it as a Python library. 

\section{Results}
\label{sec:results}

\subsection{Atmospheric case study}
\label{sec:results_atmosphere}

We undertake a case study which specifically tests the systematic impact of applying different atmospheric structure assumptions, motivated by previous choices from the literature. For this case study, we consider a `median' small exoplanet defined by the surveyed population of  confirmed exoplanet candidates which have known radii and masses less than Neptune (i.e., less than $17M_\oplus$). Under this definition, our median small exoplanet has a mass $M_p=6.3 M_\oplus$, bolometric instellation $F_\mathrm{bol}=44.0\mathrm{\,So}$, and orbits a star with effective temperature $T_\mathrm{eff}=5485\mathrm{\,K}$ (see Appendix \ref{app:exoplanets}). For this case study, since exoplanet core masses remain largely unconstrained \citep{ Zeng2019, unterborn_nominal_2023}, we also adopt a constant Earth-like core mass fraction of $f_c=0.325$ \citep{lodders_planetary_1998}. The surface radius $R_s$ is $1.6R_\oplus$ (Equation \ref{eq:interior}). These case study scenarios adopt an atmosphere with a simplified binary composition of \ch{H2} and 1.25\% \ch{H2O} (metallicity $Z_a=10^{-1}$), an atmosphere mass fraction $f_a=0.02$ ($p_s=131\mathrm{\,kbar}$), and a surface temperature of 2300\,K -- representative of substantial mantle melting \citep{Andrault_Solidusand_2011}.

We consider twelve different scenario combinations in this case study, arising from four levels of complexity for setting $T(p)$ and three methods for treating gravity in our hydrostatic integration. For $T(p)$, these are: isothermal, fully adiabatic, adiabatic with a skin-temperature stratosphere, and radiative-convective equilibrium solved using energy flux conservation. For gravity, these are: (1) $g$ held constant at $g_s$, (2) $g$ calculated as $g(r)=g_s (R_s/r)^2$, or (3) $g$ solved self-consistently with the climate calculation while accounting for gravitational attraction throughout the atmosphere (Equation~\ref{eq:grav_enclosed}).

\begin{figure*}
    \centering
    \includegraphics[width=0.94\textwidth]{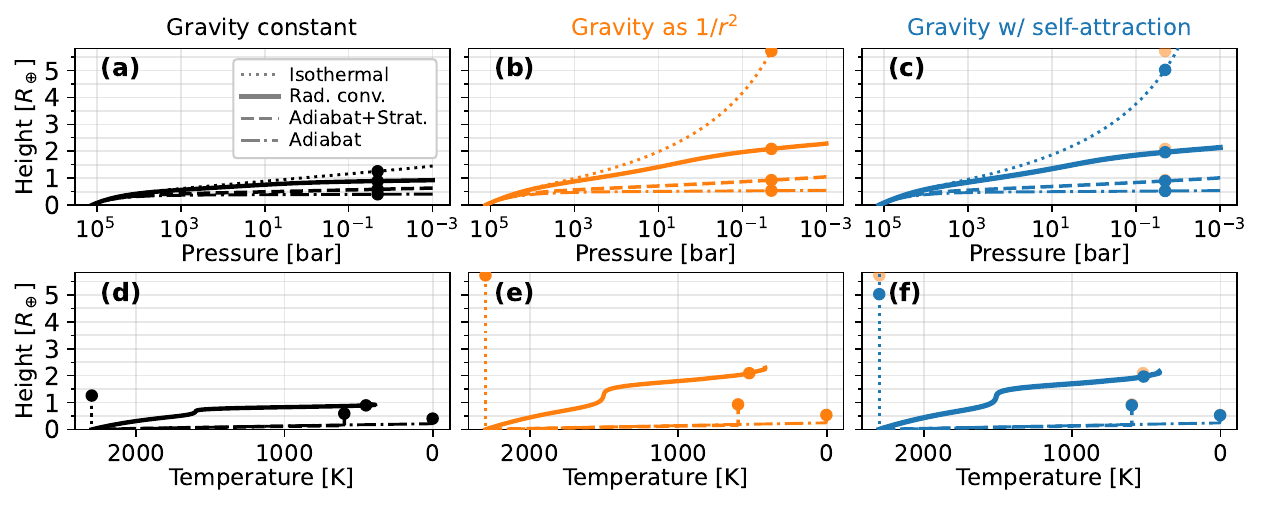}%
    \caption{Height profiles of atmospheric pressure (top row, panels \textbf{a--c}) and temperature (bottom row, panels \textbf{d--f}), calculated with AGNI for twelve combinations of scenarios: three $g(r)$ profile assumptions (line colour, columns) and four $T(p)$ profile assumptions (line style).  Photospheric radii $R_p$ are indicated by scatter points, which are carried-over between panels for the purposes of comparison. The most `realistic' modelling configuration is the solid blue line. All twelve of these modelling configurations adopt the same `median' exoplanet scenario, with a total mass of $6.3M_\oplus$, instellation flux of 44.0\,So, and stellar effective temperature of \SI{5485}{\kelvin}, as a representative of the confirmed small exoplanet population (Appendix~\ref{app:exoplanets}).
    }
    \label{fig:atmosphere}
\end{figure*}

Figure \ref{fig:atmosphere} plots the pressure and temperature profiles for these twelve scenarios as a function of height $z$ from the planet's surface: showing $p(z)$ in the top row and $T(z)$ in the bottom. Overall, the atmospheres show substantial differences in profile shape and radial extent, resulting in photospheric heights $z_p$ (scatter points) varying between 0.4~and~$5.7R_\oplus$ above the surface.

The most important differences arise from adopting different $T(p)$ assumptions (line style). Consider the $T(p)$ profiles in Figure \ref{fig:atmosphere}f: the adiabatic profile (dashdot) has the steepest lapse rate, so it results in cooler temperatures aloft which have smaller scale heights, and correspondingly produces a shallower atmosphere ($z_p\approx0.5R_\oplus$). In comparison, the isothermal profile has a zero lapse rate, so it yields a substantially hotter and more extended atmosphere $z_p\approx5.0R_\oplus$. Our energy-conserving radiative-convective treatment (solid line) yields a more complex temperature structure, and a photospheric radius intermediate to the two end-members. In the constant gravity scenarios (black lines), the range of photosphere heights which arise from our choice of $T(p)$ structure (line style) is $\approx0.8R_\oplus$. This sensitivity to $T(p)$ structure increases to a range of $\approx4.5R_\oplus$ in the more physically complete scenarios (blue lines), which derive a radially-dependent gravitational field (Equation~\ref{eq:grav_enclosed}). Both sensitivity ranges exceed the $\approx0.5 R_\oplus$ radius differences which arise from varying core mass fraction (Figure \ref{fig:interior}), highlighting that atmospheric modelling choices can -- in this representative exoplanet regime -- overprint inferences of variables controlling planetary interior structure. 

The atmosphere height also varies substantially, depending on how $g(r)$ is treated. The solid lines across the panels of Figure \ref{fig:atmosphere} all represent radiative-convective equilibrium with different gravitational treatments. Comparing the radiative-convective cases (solid lines) across the two simpler $g(r)$ treatments, the atmosphere height increases from $\approx1R_\oplus$ (panel a) to $\approx2R_\oplus$ (panel b). The difference being that gravity decreases quadratically with radius in the latter case, corresponding to a larger scale height in the upper atmosphere and larger planetary radii. Models which account for attraction of the upper-atmosphere to its deeper layers are shown in panels (c) and (f), which are somewhat less extended than the $1/r^2$ cases because of the increased gravity from the enclosed-atmosphere mass (compare orange versus blue scatter points in panel f).

In the isothermal cases where gravity varies with height (dotted lines in panels b and c), temperature-height profiles show a runaway behaviour where $z$ increases rapidly as a function of our pressure coordinate. While potentially pathological, this behaviour is expected \citep{owen_evap_2017, owen_review_2019}. Inspection of the structure equations above suggests a potential for extremely extended atmospheres: low values of $g$ generate large $\dd r/ \dd p$, which further decrease $g$ as $\dd p$ is integrated upwards in altitude. For example, a hot 2000\,K pure-\ch{H2} atmosphere on Earth would have a large scale height of 0.13 Earth radii. It is generally expected that highly-irradiated and/or extensive atmospheres should become weakly bound by gravity, which is also suggested to be the case for \PiMen\,c \citep{Munoz_pimenc_2021}. Here, all modelled atmospheres are treated as hydrostatic -- we do not simulate escape processes or Roche Lobe overflow -- although we do directly consider regimes in which upper-atmospheres become weakly bound by gravity.

These results highlight that adopting various $T(p)$ assumptions can radically shape the sizes of modelled planets, all else equal. A corollary of this sensitivity is that planetary parameters \textit{inferred} from observed radii are systematically sensitive to our method for modelling $T(p)$.

\subsection{Mass-radius relations and their physical sensitivities}
\label{sec:results_curves}

We present our library of planetary structure models which span the parameter space defined by Table~\ref{tab:params}. This parameter space encompasses planet masses from 1~to~$10 M_\oplus$, various atmospheric compositions (which could feasibly arise from diverse formation pathways and magma ocean degassing), and different irradiation regimes (from the ultra-hot to the potentially-habitable). To simulate each scenario, Equation \ref{eq:interior} is used to calculate interior radius $R_s$, and then our radiative-convective atmosphere climate model (Section \ref{sec:methods_planet}) is applied to calculate $R_p$ and other derived variables. In this subsection, and the following ones, we extract cross-sections from our grid which explore regimes that satisfy our strong convergence criterion.

\begin{figure}
    \centering
    \includegraphics[width=\linewidth]{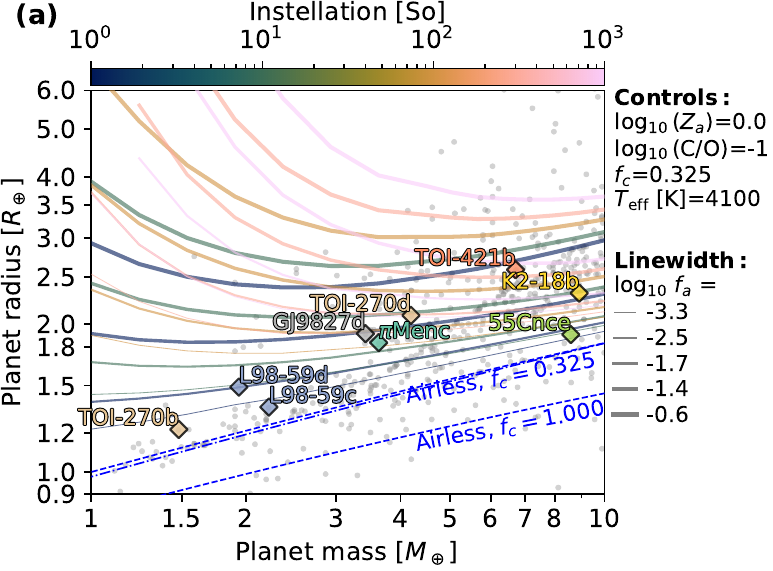}%
    \vspace{3mm}
    \includegraphics[width=\linewidth]{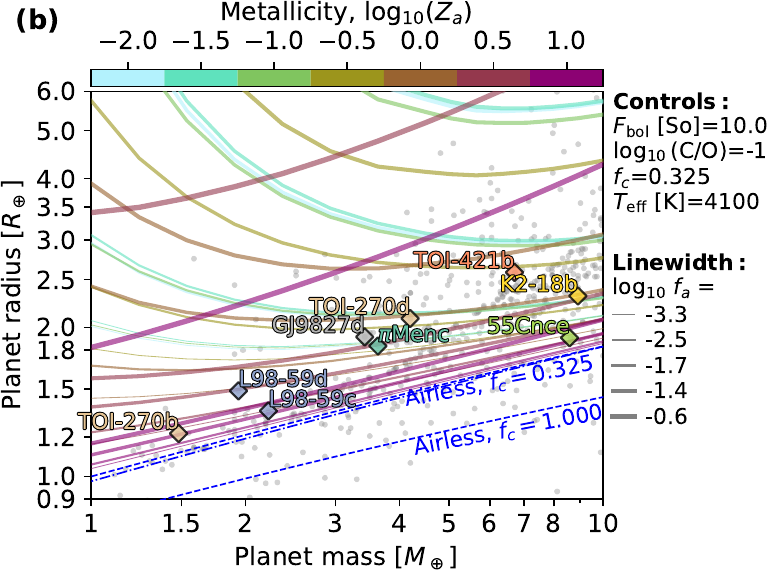}%
    \caption{
    Isolines of planet radius versus mass, extracting cross-sections from our grid of models by holding other variables constant (top-right text box). \textbf{Top (a)}: sensitivity to instellation (colour) and atmosphere mass fraction $f_a$ (line width, five values). \textbf{Bottom (b)}: sensitivity to atmospheric metallicity $\log_{10} Z_a$ (colour) and $f_a$ (line width, five values). For visualisation, only five sets of secondary variables (line widths) are shown. Dashed blue lines show airless cases with different core sizes \citep{Zeng2019}. The dash-dotted blue line shows the airless $f_{ci}=0.325$ case calculated using Equation~\ref{eq:interior}. Scatter points show the masses and radii of  6260 `confirmed' exoplanets using data obtained from \url{exoplanet.eu}  on 27\,February\,2026.
    }
    \label{fig:mr}
\end{figure}

Figure \ref{fig:mr} shows two subsets of mass-radius isolines, holding different grid parameters constant in each case. We plot these mass-radius lines alongside the currently-surveyed population of exoplanets (scatter points), with data obtained from \url{exoplanet.eu}. Some exoplanets appear at radii below the dashed-blue airless lines because of their observational uncertainties (error bars not shown).

The coloured lines in the top panel (a) of Figure \ref{fig:mr} correspond to different bolometric instellation fluxes (colours) and atmosphere mass fractions (line widths). Immediately, it is clear that the planet mass-radius relationship is sensitive to both of these parameters, all else equal. There is an overall trend of $R_p$ increasing with instellation, caused by lower scale heights and lower gas densities at the higher planet temperatures, even at this $74\times$ super-Solar metallicity. Given substantial variations between instellation scenarios (line colours), $F_\mathrm{bol}$ is clearly a strong determiner of atmospheric structure, particularly at lower masses. For example, the mass-radius coordinates of \PiMen\,c or GJ\,9827\,d are consistent with several isolines on the plot, and could therefore be taken to result from a range of atmospheric mass fractions, depending on the instellation considered. Instellation fluxes must be considered when calculating planetary structures, and when comparing generalised mass-radius curves to the observed radii of \textit{specific} exoplanets. These degeneracies are further exemplified by the sub-Neptune TOI-270\,b, which is incompatible with all instellations and $f_a$ shown in panel (a); more substantial metal-enrichment than $Z_a=1$ would be required to explain its structure \citep{benneke_jwst_2024, teske_metal-rich_2019}. JWST measurements of TOI-270\,b's atmospheric composition have found evidence of \ch{CH4}+\ch{CO2}+\ch{H2O} chemistry and inferred a metal mass fraction of 58\% \citep{benneke_jwst_2024}. Given this planet's irradiation exposure, it has likely been subjected to thermal escape processes throughout its lifetime that could efficiently remove volatiles from its atmosphere, which points to metal-enrichment during its formation process \citep{ji_the_2025, cherubim_oxidation_2025}.

The bottom panel (b) of Figure \ref{fig:mr} instead takes $Z_a$ as the discriminating variable, keeping instellation flux constant at 10\,So. The highest metallicity cases (purple lines) strongly overlap with each other and tend towards the airless scenarios: a well-known degeneracy which results from high molecular weights and atmospheres of smaller vertical extents \citep{barstow_outstanding_2020, madhusudhan_atmospheric_2018}. Isoline convergence towards the airless $f_{ci}=0.325$ structures from \citet{Zeng2019} (dashed blue) and calculated from Equation~\ref{eq:interior} (dash-dotted blue) validates the behaviour of our coupled interior-atmosphere structure in the small-$f_a$ end-member regime. 

It remains possible for high-metallicity scenarios to still become rather extended beyond the airless scenario when the atmosphere mass fraction is large. For example, the mass-radius of L\,98-59\,d sits on a purple $Z_a=10$ line; a volatile rich scenario for this super-Earth has been previously justified by planetary evolution modelling and evidenced by detections of \ch{H2S} in its atmosphere \citep{nicholls_escape_2025, banerjee_atmospheric_2024, cheverall_ground_2026}. Low-metallicity scenarios (blue and green lines) tend to cluster together because these all correspond to compositions dominated by \ch{H2} (Figure~\ref{fig:chem}). 

\begin{figure}
    \centering
    \includegraphics[width=\linewidth]{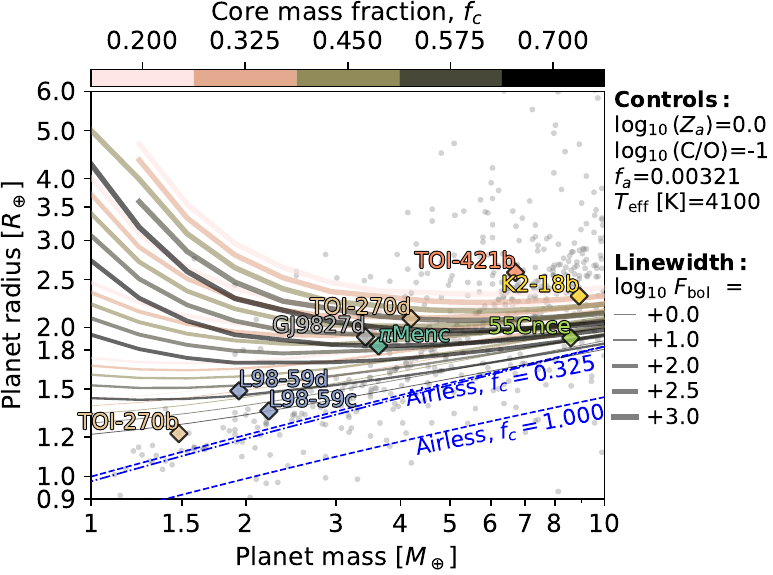}%
    \caption{
    Radius versus mass, as functions of metallic-core mass fraction (colour) and instellation (line width). Other grid variables held constant. Dashed blue lines show airless cases with different core sizes \citep{Zeng2019}. The dash-dotted blue line shows the airless $f_{ci}=0.325$ case calculated using Equation~\ref{eq:interior}. Scatter points show the masses and radii of  6260 `confirmed' exoplanets using data obtained from \url{exoplanet.eu}  on 27\,February\,2026.
    }
    \label{fig:mr_c}
\end{figure}

Similarly, Figure~\ref{fig:mr_c} compares structural sensitivity to metallic-core mass fraction $f_c$ and instellation flux. These variables largely counter each other, highlighting that retrievals of core sizes from the observed mass-radius combinations of enveloped planets must incorporate atmosphere models which are sensitive to planetary irradiation environments. Degeneracy between these parameters is weaker for lower-mass planets where low surface gravities and absorbed stellar radiation can readily inflate their atmospheres.

\subsection{Gravitational structure when vertically resolved}
\label{sec:results_gravity}

We have modelled a range of atmosphere mass fractions $f_a$, including scenarios where the atmosphere comprises a substantial fraction of the total planet mass. It is expected from Newton's Shell Theorem that gravity $g(r)=GM(r)/r^2$ varies as a function of radial distance $r$. A component of this function declines for larger radial distances ($\propto1/r^2$), which is partially offset as the $\propto M(r)$ component increases with $r$ as more mass is enclosed within each atmosphere layer. Additional downward attraction from a growing $M(r)$ component will act to increase the photospheric gravity $g_p$ compared to the first-order $1/r^2$ scaling, and thus decrease the scale height $H_p$ of the atmosphere probed by transmission spectroscopy. Potential outcomes from the competition between these components are demonstrated in Figure~\ref{fig:atmosphere}. Now, we empirically compare the outcome of these competing effects across different atmosphere mass fraction regimes.

\begin{figure}
    \centering
    \includegraphics[width=\linewidth]{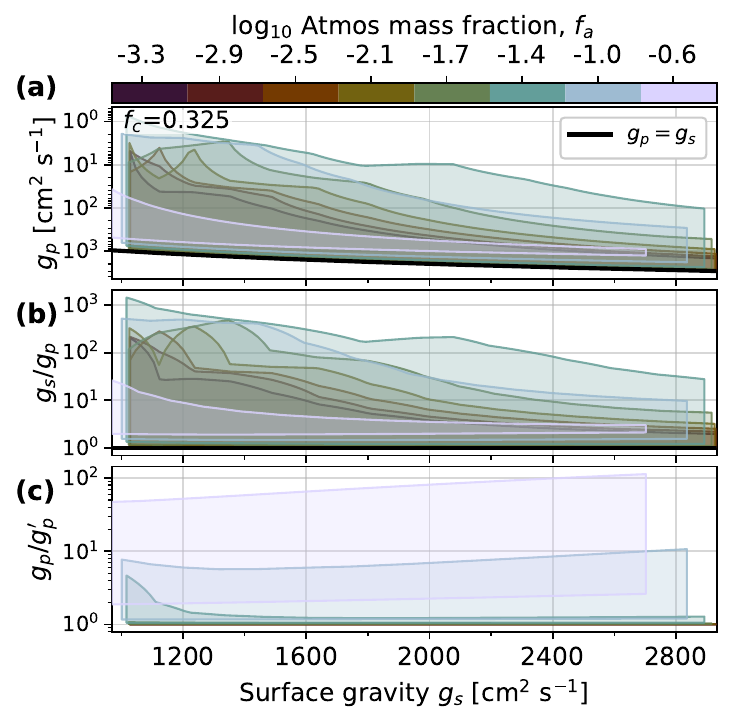}%
    \caption{
     Three comparisons of the empirical height-variations in gravitational acceleration, showing grid cases with metallic-core mass fractions $f_c=0.325$. Each shaded `envelope' plots the full min/max range of the y-axis quantities, for a given atmosphere mass fraction (colour, $f_a$), for the corresponding value of surface gravity $g_s$. The x-axis is a proxy for the total planet mass, independent of $f_a$. \textbf{Top (a)}: the range of photosphere gravities $g_p$ at each $g_s$ and $f_a$. \textbf{Middle (b)}: ratio of surface gravity to that at the photosphere $g_s/g_p$, for each $g_s$ and $f_a$. \textbf{Bottom (c)}: ratio of the self-consistently calculated photosphere gravity $g_p$ to the simpler scaling of $g_p'=g_s(R_s/R_p)^2$, which does not account for self-attraction ($g_p'$) -- comparable to Figure~\ref{fig:atmosphere}b. The black line demarcates the constant-gravity limit of $g_p=g_s$.  
    }
    \label{fig:gravity}
\end{figure}

The top panel (a) of Figure~\ref{fig:gravity} plots the photospheric gravity $g_p$ versus surface gravity $g_s$ for all strongly-converged cases with core fractions $f_c=0.325$. The x-axis serves as coordinate of planet mass which is independent of $f_a$. Shaded regions are coloured by atmosphere mass fraction $f_a$, which all lie above the black $g_p=g_s$ line because gravity at the photosphere is less than that at the surface. However, noting the logarithmically scaled y-axes, photospheric gravity can decrease to values below $\sim\sfrac{1}{10}^\mathrm{th}$ or even $\sim\sfrac{1}{100}^\mathrm{th}$ of the surface gravity$g_s$, highlighting the importance of considering a radially-variant gravity. 

To identify these specific  behaviours, the middle panel (b) of Figure~\ref{fig:gravity} plots the ratio of $g_s/g_p$. Deviation from the $g_p=g_s$ end-member is not a linear function of $f_a$ because of competition between the $M(r)$ and $1/r^2$ components. The most massive $f_a$ generally tend closer to the $g_p=g_s$ regime because the additional mass of vapour encompassed by the photosphere leads to $M(R_p)$ being substantially larger than $M(R_s)$, outcompeting the effect of $R_p>>R_s$. On the other hand, some less-massive atmospheres also tend closely towards the $g_p=g_s$ limit because their surface pressures are small, so $R_p/R_s\approx 1$. 

The bottom panel of Figure~\ref{fig:gravity} compares the photospheric gravity $g_p$ to a first-order radial scaling of the surface gravity: $g_p'=g_s\times(R_s/R_p)^2$, neglecting the component of atmosphere mass on the photosphere. Deviation of this ratio above unity is a direct probe of the $M(r)$ component across the hydrostatic atmosphere. In cases with large envelopes, the contribution of the atmosphere mass to the total mass is significant enough that the actual photospheric gravity is up to 100 times that predicted by first-order $1/r^2$ scaling. From an observational perspective, where a planet's radius has been measured from its transit depth, neglecting this $M(r)$ contribution would lead to a systematically under-estimated atmosphere mass fraction $f_a$ or an over-estimated surface gravity.

\subsection{Atmospheric convection versus instellation and composition}
\label{sec:results_convect}

The presence and strength of atmospheric convection is important because it determines temperature structure, energy transport, and whether chemical abundances are homogeneously quenched by its efficient mixing. It is not \textit{a priori} obvious which regimes are expected to be convectively unstable. Our radiative-convective calculations naturally permit an investigation into the empirical trends of convective vigour and location, as functions of the observationally-accessible parameters in our grid of models.

Convection is caused by buoyancy forces overcoming gravitational attraction. Buoyancy may be triggered by a decrease in the density of a fluid parcel (e.g., when heated by absorption of radiation) and potentially stabilised by compositional stratification \citep{guillot_condensatio_1995, pierrehumbert_book_2010, habib_convection_2024}.  Our modelling adopts a mixing-length parametrisation of convection, accounting for the combined effects of incoming stellar radiation, planet size, and atmospheric composition. Here, we proxy convective vigour by quantifying the vertical-maximum value of eddy diffusion coefficient $K_{zz}$ calculated by our convection parametrisation, for each grid point in our library of models. 

\begin{figure}
    \centering
    \includegraphics[width=0.94\linewidth]{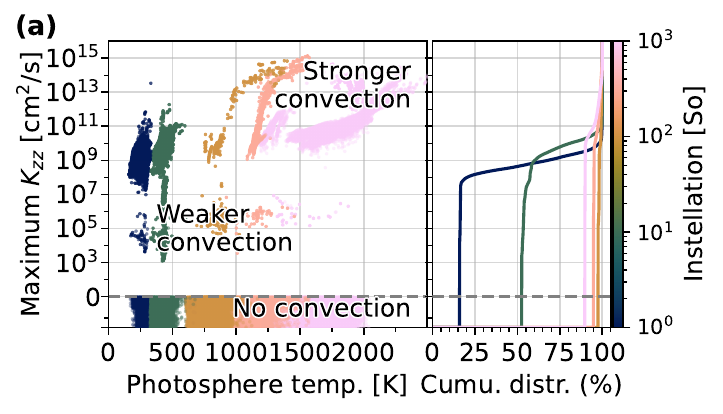}%
    \vspace*{-2mm}
    \includegraphics[width=0.94\linewidth]{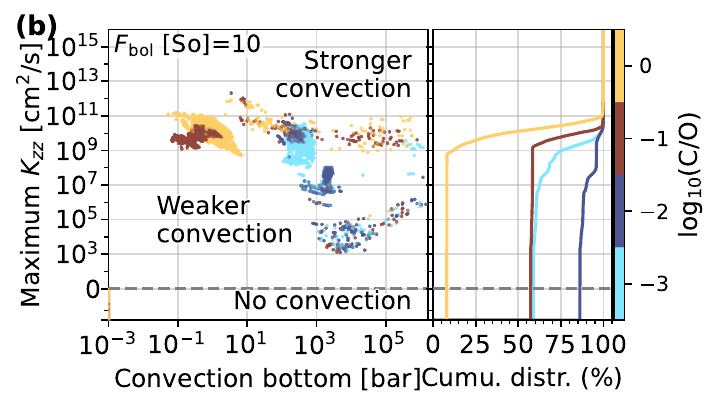}%
    \caption{
    Two projections of atmospheric convective vigour from our grid of structure models, quantified by the maximum eddy diffusion coefficient \Kzz within each grid point's atmosphere column. \textbf{Top (a)}: convective vigour \Kzz versus photospheric temperature, as functions of bolometric instellation $F_\text{bol}$ (marker colour). \textbf{Bottom (b)}: convective vigour \Kzz versus pressure-space location, as functions of atmospheric C/O ratio (marker colour), for $F_\text{bol}=\SI{10}{\soli}$ only.
    }
    \label{fig:convection}
\end{figure}

The top panel of Figure~\ref{fig:convection} plots $K_{zz}$ versus photospheric temperature $T_p$, as a function of bolometric instellation flux (colourbar). The trend between photospheric temperature (x-axis) and bolometric flux is largely monotonic, as expected, since increasing irradiation efficiently heats these planets' upper atmospheres, largely independently of their specific compositions \citep{pierrehumbert_book_2010, guillot_2010}. We might also expect that highly-irradiated regimes are convective, yet we find that most of the scenarios we have considered are not convective at all. Energy is instead carried entirely by radiative processes -- even in the more highly-irradiated regimes \citep{janssen_sulfur_2023, piette_rocky_2023, selsis_cool_2023}. These statistics are quantified by the cumulative distribution functions in the rightmost panels the plot. Instead, absorption of radiation by molecular continua (particularly \ch{H2}-\ch{H2} collisional absorption) is efficient at heating the atmosphere to high temperatures and increasing its radiative diffusivity, allowing energy to be efficiently carried by radiative diffusion, rather than by convection. For a given instellation flux, the calculated $K_{zz}$ range eight orders of magnitude due to variations in atmosphere gravity and composition (e.g., $F_\mathrm{bol}=10$, corresponding to green points in Figure~\ref{fig:convection}a). Whether convection occurs within the atmospheres of these rocky exoplanet scenarios is dependent on their composition, in addition to their instellation flux.

To focus on these latter effects, the bottom panel of Figure~\ref{fig:convection} keeps instellation fixed at $F_\mathrm{bol}=10\mathrm{\,So}$. Instead, we plot the maximum $K_{zz}$ against the convective region location, as functions of the atmosphere C/O ratio (colourbar). Overall, most scenarios remain fully-radiative, but there are clear structures in the relationship between $K_{zz}$ and $p_\mathrm{conv}^\mathrm{bot}$: notably, a narrow band of cases with $K_{zz}$ around \SI{e10}{\cmsps} and a more complex non-linear behaviour elsewhere. The grid points simulated at the largest C/O ratios are more likely to be convective than not (yellow cumulative distributions), and primarily populate the narrow range of large $K_{zz}$ values. 

The trends in Figure~\ref{fig:convection}b are an outcome of the different processes which trigger the convection in these atmospheres. At high C/O (yellow points), the convection is only triggered by absorbing stellar radiation -- yielding a narrow range of large $K_{zz}\approx\SI{e10}{\cmsps}$. These yellow-coloured cases are joined by a subset of low-C/O grid points (blue colour), which also sit within the $K_{zz}=\SI{e10}{\cmsps}$ band, and also correspond to climates where stellar radiation is the primary driver of convection. Meanwhile, a separate subset of the low-C/O grid points (blue) show a non-linear relationship between $K_{zz}$ and $p_\mathrm{conv}^\mathrm{bot}$, at smaller $K_{zz}$. Although we note once-more that most low-C/O cases are non-convective, the low-but-finite-$K_{zz}$ subset can be explained by the thermochemical production of \ch{CO2} and \ch{H2O} in metal-rich atmospheres, towards smaller C/O ratios (Figure~\ref{fig:chem}). Formation of \ch{CO2} and \ch{H2O} increases the opacity of the deep atmosphere, which allows the (small) internal heat production to trigger deep convection in cases that would otherwise be fully-radiative  \citep{cmiel_structure_2025, nicholls_convective_2025}. The shape of the complex $K_{zz}$-$p_\mathrm{conv}^\mathrm{bot}$ relationship across these lower-$K_{zz}$ cases (blue points) is a result of $K_{zz}$ being a linear function of gravity, which Section~\ref{sec:results_gravity} demonstrated to be highly sensitive to the particular climate-structure scenario considered. The vertical stratification of mean molecular weight is minor in all cases because we model these atmospheres as being elementally well-mixed. Overall, these behaviours highlight another important consequence of a self-consistent treatment of both climate and gravity.

Convectively unstable regions also typically have larger lapse rates than radiative layers \citep{peng_puffy_2024, nicholls_convective_2025}. When present, atmospheric convection can generate substantial temperature differences between the observable upper atmosphere and the deeper regions of a planet. So, in addition to its control over compositional homogeneity, atmospheric temperature-pressure profiles are fundamentally related to the thermal conditions at a planet's surface. In the next section, we empirically consider the range of surface conditions expected for exoplanets hosting substantial atmospheres. 

\subsection{Parallel interpretative frameworks: mass-radius and the habitable zone }
\label{sec:results_hz}

Which irradiation and compositional scenarios allow for potentially habitable conditions at the surface, and in which scenarios is the surface is more likely to be molten? The former question is classically approached by appealing to the habitable zone concept: a region of $F_\mathrm{bol}-T_\mathrm{eff}$ phase space in which an approximately Earth-sized planet could potentially sustain liquid water on its surface through various climate feedbacks \citep{kopparapu_habitable_2013, kasting_hz_1993}.  \citet{calder_most_2025} used numerical simulations of sub-Neptune evolution to explore which sub-Neptunes may host permanent magma oceans. Their simulations found that melting within sub-Neptune interiors strongly depends on atmospheric mass fraction, due to the effects of atmospheric blanketing, even for planets orbiting within their star's canonical habitable zone. In the spirit of the habitable zone and the cosmic shoreline \citep{zahnle_cosmic_2017} concepts, \citet{calder_most_2025} then proposed the `solidification shoreline' concept to predictively map the transition between molten and solidified interiors. Various shoreline boundaries were defined in $F_\mathrm{bol}-T_\mathrm{eff}$ phase space, depending on atmosphere mass fraction $f_a$. These first solidification shoreline maps were limited to low metallicity  gas-dwarf planets \citep{ginzburg_super-earth_2016, valencia_diversity_2025}. Here, we explore which scenarios in our compositionally diverse model grid correspond to clement, supercritical, and molten surficial environments.

\begin{figure}
    \centering
    \includegraphics[width=0.92\linewidth]{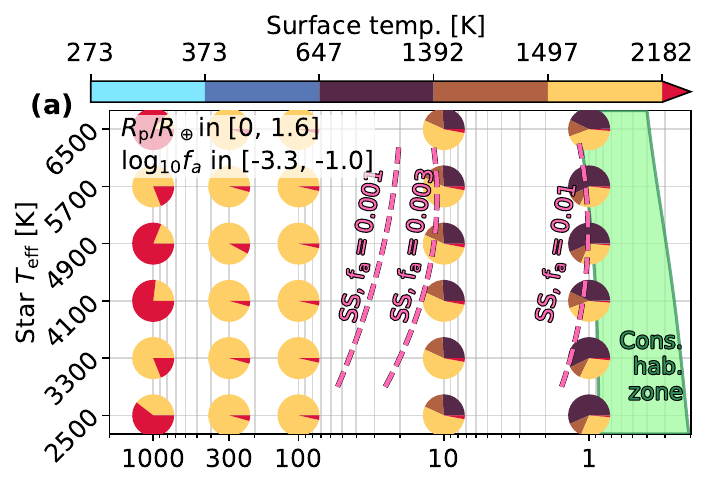}%
    \vspace*{-2mm}
    \includegraphics[width=0.92\linewidth]{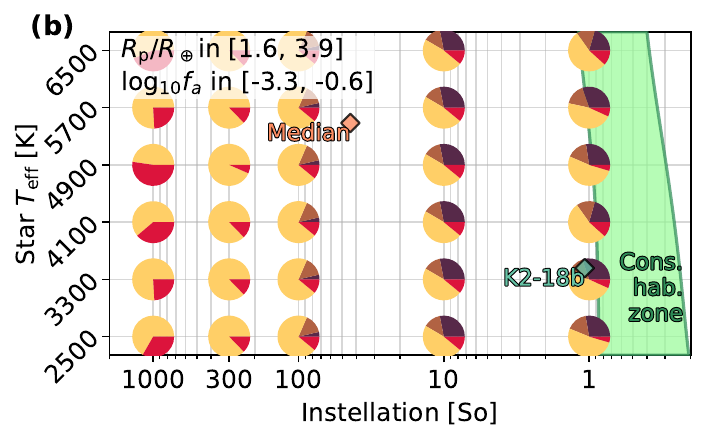}%
    \caption{
    Modelled surface temperatures $T_s$ as a function of instellation flux (x-axis) and stellar effective temperature (y-axis). \textbf{Top:} Earth-sized and super-Earth planets. \textbf{Bottom:} sub-Neptune planets. Colourbar boundaries correspond to $T_s$ of \ch{H2O} solidification at 1 bar, \ch{H2O} condensation at 1 bar, \ch{H2O} supercriticality (647\,Kelvin), and \ch{MgSiO3} solidification at $10^3$~bar, $10^4$~bar, and $10^5$~bar \citep{wolf_equation_2018}. Only 78 grid points have sub-critical surface conditions (light and dark blue colours); they comprise only small pie-chart sectors, and are discussed in detail within Section~\ref{sec:results_hz}. The green scatter point specifically highlights the sub-Neptune K2-18\,b,and the blue scatter point indicates a `median' small exoplanet as per Appendix~\ref{app:exoplanets}, using data drawn from \url{exoplanet.eu}. The conservative habitable zone is shown in green \citep{kopparapu_habitable_2013}. Cross-sections of the `solidification shoreline' are shown as dashed pink lines, as a function of atmosphere mass fraction $f_a$ \citep{calder_most_2025}.
    }
    \label{fig:hz}
\end{figure}

Figure~\ref{fig:hz} plots the distribution of modelled surface temperatures (colourbar) as pie charts, for each $F_\mathrm{bol}-T_\mathrm{eff}$ combination in our grid of models. The conservative habitable zone is indicated by the green region \citep{kopparapu_habitable_2013}, and various solidification shorelines (dependent on $f_a$) by the dashed pink lines. As expected, our climate simulations (pie charts) yield smaller surface temperatures at lower instellations, yet very few scenarios correspond to potentially habitable conditions. 

Across all modelled scenarios, only 78 grid points have surface temperatures below the critical point of water (647~Kelvin).  In the combined terrestrial-sized and super-Earth regime (panel a), the subset of 48  sub-critical scenarios all correspond to our lowest instellation flux of $F_\mathrm{bol} = 1\mathrm{\,So}$;  larger instellations (i.e., $F_\mathrm{bol} \ge 10$\,So) yield super-critical surface temperatures.  Just 30 scenarios have sub-critical surface temperatures in the sub-Neptune regime (panel b), which is reflective of large atmosphere mass fractions that induce a strong greenhouse effect \citep{pierrehumbert_hydrogen_2011,  krissansen_erosion_2024, boer_absence_2025}.  We identify no clear trend between surface supercriticality and stellar effective temperature; more simulations are necessary to robustly identify a trend from the statistics in this part of parameter space \citep{kopparapu_habitable_2017, wolf_constraints_2017}. The irradiation environment which yields the largest fraction of subcritical outcomes is $F_\mathrm{bol}=1$\,So and $T_\mathrm{eff}=4100$\,K -- a cooler photospheric temperature than the Sun \citep{kopparapu_habitable_2013, gueymard_suns_2004}.

The sub-Neptune exoplanet K2-18\,b ($T_\mathrm{eff}=3457\,K$, $F_\mathrm{bol}=1.2\mathrm{\,So}$), indicated in Figure~\ref{fig:hz}b, has a substantial \ch{H2}-dominated atmosphere. The magma ocean scenario has been previously suggested for K2-18\,b, which would be consistent with the non-detections of \ch{NH3} in this planet's atmosphere \citep{shorttle_k218b_2024}. On the basis of our climate models, it is apparent from its corresponding pie chart that a sub-Neptune subjected to this irradiation environment can readily maintain surface temperatures beyond the critical point of water, which is incompatible with a surface ocean of liquid water. Many exoplanets residing within the habitable zone may also host deep magma oceans. For our scenarios proximate to the habitable zone, $F_\mathrm{bol}=1\mathrm{\,So}$ scenarios, with thinner $f_a=10^{-3.3}$ atmosphere mass fractions, only 0.6\% of cases have surface temperatures above their corresponding pressure-dependent \ch{MgSiO3} solidus temperature, implying solidified interiors, since their atmospheres do not induce sufficient greenhouse heating. However,  for larger atmospheres $f_a=10^{-2.5}$, a 31.7\% of our scenarios do maintain deep magma oceans in the habitable zone,  at $F_\mathrm{bol}=1.0\mathrm{\,So}$. Blanketing and greenhouse heating effects of these atmospheres -- dependent on their mass and composition -- acts in competition with the `overburden' pressure they apply to their interiors, which increases the solidus temperature  \citep{Andrault_Solidusand_2011, wolf_equation_2018, breza_not_2025}. These outcomes can be identified through a unified inference of whole-planet structures (Section~\ref{sec:methods_infer}) with specific chemical markers, such as \ch{NH3}  depletion \citep{ortenzi_mantle_2020,shorttle_k218b_2024, nicholls_escape_2025}.

Large surface pressures can, in principle, stabilise planetary interiors against substantial melting \citep{breza_not_2025}. Yet, we find that a large fraction of our canonically-temperate scenarios instead reside in a molten regime (scarlet pie slices) -- correspondent with the $f_a=0.01$ solidification shoreline -- even when accounting for this pressure effect. Under our more strongly irradiated scenarios (left side of Figure~\ref{fig:hz}), the calculated surface temperatures are typically in excess of the \ch{MgSiO3} solidus temperature \citep{wolf_vaporock_2023}. The `median' small exoplanet (as defined in Appendix~\ref{app:exoplanets}) has an instellation flux of 44~So -- which represents a greater exposure to stellar radiation than even the most conservative solidification shoreline (pink line) and sits within the domain of our largely-molten scenarios (pie charts). That the `typical' small exoplanet sits within the molten regime emphasises the growing opportunity for probing deep planetary interiors through large-scale exoplanet surveys, enabled by characterisations of their atmospheric compositions \citep{lichtenberg_constrainin_2025}. 

We predict very few habitable planet scenarios, yet these environments must exist within some part of parameter space -- evidenced by the Earth, at least. We can interpret this result as habitable surface conditions requiring smaller atmosphere mass fractions (surface pressures) than those generally explored by our grid of models. Otherwise, extremely high shortwave albedos are necessary to contrive habitable surface conditions on sub-Neptune planets \citep{barrier_general_2025,jordan_planetary_2025}. Alternatively, we can interpret this result as an extension of previous suggestions that a strong hydrogen greenhouse effect can shift the habitable zone to much larger orbital distances with lower instellations \citep{pierrehumbert_hydrogen_2011, seager_laboratory_2020}.

Section~\ref{sec:results_curves} demonstrated that attempts at direct comparisons between theoretical mass-radius isolines and the measured mass-radius combination of a \textit{specific} exoplanet is difficult and vulnerable to physical systematics. In this section, we have demonstrated that similar difficulties are encountered when comparing super-Earth and sub-Neptune planets to habitable zone boundaries. Instead, the very same structure models used to calculate planetary structure and mass-radius curves can be applied to estimate `hidden' parameters of interest (e.g., $T_\mathrm{surf}$), while remaining sensitive to measurement uncertainties and physical systematics.

\subsection{Demonstrations retrieving whole-planet structures}
\label{sec:results_ia}

Our library of radiative-convective planetary structure models provides a rich basis for performing retrievals to estimate the parameters of \textit{particular} exoplanets. In contrast to a `free' retrieval, the physics in these models will limit the parameters to physically-feasible regimes while making predictions for other (hidden) variables.

Following the methodology described in Section~\ref{sec:methods_infer}, we apply our Python-based MCMC Bayesian inference tool under two demonstration scenarios. Firstly to TOI-421\,b, which is suggested to have a substantial atmosphere of near-Solar metallicity free of aerosols (Section~\ref{sec:results_ia_demo}). Secondly, to \PiMen\,c, for which detections of escaping carbon atoms and non-detection of Lyman-$\alpha$ absorption are suggestive of a hydrogen-poor composition (Section~\ref{sec:results_ia_Fbol}).

\subsubsection{Validation of retrieval framework -- TOI-421\,b}
\label{sec:results_ia_demo}

TOI-421\,b is a warm sub-Neptune exoplanet which receives a bolometric instellation of $176\pm7\mathrm{\,So}$ from its G-type host star ($T_\mathrm{eff}=5219\pm64\mathrm{\,K}$). This planet has a mass $6.7\pm0.6 M_\oplus$ and radius $2.67\pm0.08 R_\oplus$ , so its corresponding bulk density necessitates a substantial inventory of volatiles \citep{krenn_toi421b_2024}. Recent JWST NIRISS and NIRSpec observations suggest that its atmosphere is primarily composed of \ch{H2}, while ruling-out a metal rich composition \citep{davenport_toi421b_2025}. The JWST transmission spectra provide statistical evidence for \ch{H2O}, and hint at the presence of \ch{CO} and \ch{SO2} at low statistical significance. Previous application of spectroscopic retrievals to these spectroscopic measurements have estimated $\mu_p = 2.66^{+3.6}_{-0.3} \mathrm{\SI{}{\amu}}$ \citep{davenport_toi421b_2025}. 

Here, we perform a retrieval on the whole-planet structure of TOI-421\,b as a representative of the gas-dwarf sub-Neptune population. The retrieval was run for 9000 steps with 24 walkers, resulting in a runtime of 152 seconds on an AMD Ryzen laptop using 15 CPU cores. The observables used to constrain the model have uniform priors: $R_p$, $M_p$, $T_\mathrm{eff}$, $F_\mathrm{bol}$, $\mu_p$.  

\begin{figure}
    \centering
    \includegraphics[width=\linewidth]{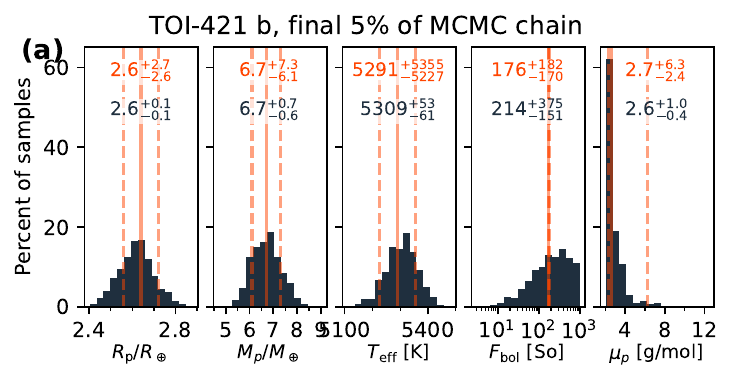}%
    \vspace*{0mm}
    \includegraphics[width=\linewidth]{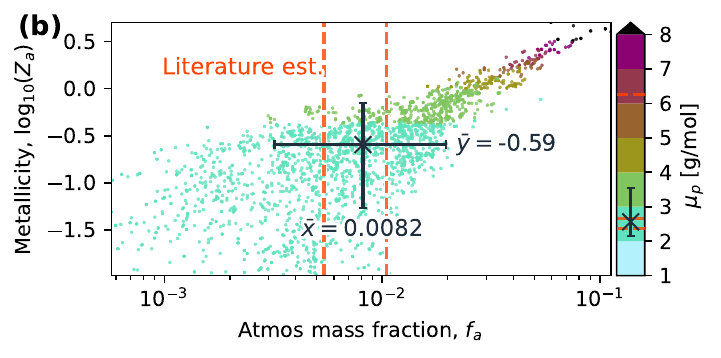}%
    \vspace*{0mm}
    \includegraphics[width=\linewidth]{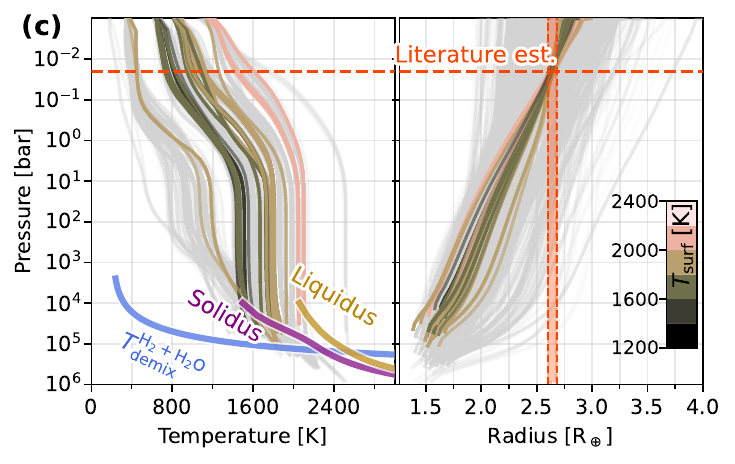}%
    \caption{
    Retrieval on TOI-421\,b showing the final 5\% of samples from the MCMC chain, compared to estimates $\pm1\sigma$ on observables used to constrain the retrieval (orange lines; \citet{davenport_toi421b_2025}). \textbf{Top (a)}: histograms of observables calculated by the forward model (black). \textbf{Middle (b)}: atmospheric metallicity $\log Z_a$, mass fraction $f_a$, and molecular weight $\mu_p$ sampled by the retrieval, with their median values $\pm1\sigma$ indicated by the black points. \textbf{Bottom (c)}: atmosphere temperature $T(p)$ and radius $r(p)$ profiles for the MCMC samples. Well-fitting cases with radius errors <1\% are shaded by their surface temperature (colourbar), or highlighted in blue if they cross the \ch{H2}-\ch{H2O} demixing binodal. Poorer-fitting cases are shaded in grey. Over-plotted are the \ch{H2}-\ch{H2O} demixing binodal derived from a combination of \textit{ab-initio}  density functional theory and molecular dynamics simulations \citep{howard_the_2025}, and the pure-\ch{MgSiO3} solidus and liquidus melting curves \citep{wolf_equation_2018}.
    }
    \label{fig:ia_demo}
\end{figure}

Figure~\ref{fig:ia_demo} plots the final 5\% of samples from the MCMC chain, flattened over all walkers. The top panel (a) plots posterior distributions from the sampled observable variables, with the `true' observed values indicated by the vertical orange lines: median (solid), $\pm 1 \sigma$ (dashed). For all five variables, the median of the MCMC samples (black text) converges upon the truth (red text) within these 9000 steps. While instellation appears to exhibit the largest deviation, this difference is a reflection of $F_\mathrm{bol}$ varying on a logarithmic scale (x-axis). The molecular weight histogram has a positive tail because of the relatively large asymmetric uncertainty on the truth value adopted \citep{davenport_toi421b_2025}. Overall, these five histograms confirm that our retrieval tool can reproduce the adopted measurements of TOI-421\,b whilst utilising radiative-convective planetary structure models, at minimal computational cost. This performance is enabled by our forward model simply being an interpolator fitted to the grid of models presented in Section~\ref{sec:results_curves}.

The middle panel (Figure~\ref{fig:ia_demo}b) plots values of atmospheric metallicity $Z_a$, mass fraction $f_a$, and photospheric molecular weight $\mu_p$ from the same 5\% of MCMC samples. The horizontal orange lines on the colourbar show the $\mu_p$ estimated from JWST spectroscopy. The dashed vertical orange lines are the $\pm1\sigma$ range ($f_c=[0.54\%, 1.05\%]$) on the atmosphere mass fraction estimated by \citet{davenport_toi421b_2025} from combining calculations from an internal structure code with the HELIOS atmosphere model, which they performed separately from their retrieval. The median result from our calculations (black scatter point) is consistent with the atmosphere mass fraction inferred by \citet{davenport_toi421b_2025}: we find $f_a=0.82^{+1.12}_{-0.50}\%$, representative of a substantial volatile atmosphere and consistent with the planet's known low density. Previously, our Figure~\ref{fig:mr}b highlighted that atmosphere metallicity is largely inversely degenerate with the atmosphere mass fraction for a given mass and radius. We also infer $\log_{10}Z_a=-0.59^{+0.44}_{-0.67}$, which represents an atmospheric composition of $19\times$~solar metallicity. It is difficult to estimate $Z_a$ with small errors in this low-metallicity regime, because a wide range of scenarios lead to \ch{H2}-dominated atmospheres (Figure~\ref{fig:chem}). Metallic-core mass fraction and C/O are not well constrained by our retrieval and have posteriors representative of their uniform priors. With additional observations, the inference of a relatively low atmospheric C/O ratio would suggest a history of muted escape from this planet or be a tracer of deep magma ocean interactions which shape its atmospheric composition \citep{shorttle_k218b_2024, nicholls_redox_2024, werlen_atmospheric_2025}. 

For the purposes of our retrieval methodology, our grid of models is represented using a multidimensional linear interpolation function. In principle, this function adopts our `weak' convergence criterion on radiative-convective equilibrium (Section~\ref{sec:methods_params}). However, we note that the regimes probed by the MCMC sampler correspond to grid-points which are fully converged under our `strong' convergence criterion. Typically, only the more extreme, end-member scenarios spanned by our grid violate our strong criterion, so they are naturally avoided by the MCMC sampler algorithm. 

Our forward-model is founded upon precomputed structure calculations, so an estimation of the corresponding climate states are directly accessible from the retrieval. For each of the final 5\% of MCMC samples, we locate the nearest modelled grid points in our parameter space and extract the corresponding atmosphere profile. These pressure-temperature and pressure-radius profiles are plotted in Figure~\ref{fig:ia_demo}c, where line colours nominally indicate their surface temperature (colourbar). The subset of cases which could potentially de-mix due to \ch{H2}-\ch{H2O} immiscibility have their $T(p)$ profiles plotted in blue. A comparison between the modelled atmosphere profiles and the planet's measured radius is visualised on the right subplot: numerous $r(p)$ profiles fall within the observational uncertainty, so a modest range of $T(p)$ climate profiles (left subplot) are also feasible. 

The $T(p)$ profiles in Figure~\ref{fig:ia_demo}c broadly fall between the \ch{MgSiO3} solidus and liquidus temperature boundaries (purple and yellow lines), suggestive of this planet hosting a magma ocean. We adopt a conservative estimate on internal heat production, so the surface temperatures in these models represent lower-limits \citep{barr_tides_2018, nicholls_tidal_2025, herath_thermal_2024}. The surface pressures arising from the retrieved $f_a$ do not generally cross the \ch{H2}-\ch{H2O} demixing binodal \citep{howard_the_2025}. None of the $T(p)$ profiles representative of the best fitting MCMC samples cross the \ch{H2}-\ch{H2O} binodal curve \citep{howard_the_2025}, suggesting that this planet's atmosphere is likely to be fully miscible, which rules against the potential for a substantial immiscible \ch{H2O} inventory hidden deep within its atmosphere \citep{piaulet_water_2025}. Instead, TOI-421\,b's reducing atmosphere could exist in a well-mixed state which is maintained at thermochemical-solubility equilibrium with a deep internal magma ocean. 

\subsubsection{Constraints from physical systematics -- \PiMen\,c}
\label{sec:results_ia_Fbol}

The case of TOI-421\,b showed that compositional measurements can partially lift the degeneracy $Z_a$ and $f_a$ encountered when explaining the structure of sub-Neptune planets (Figure~\ref{fig:ia_demo}b). However, incoming stellar radiation can also inflate planetary radii by heating their atmospheres, dependent on the gas composition (Figure~\ref{fig:mr}). Given the physics relating these variables, here we compare the systematic role of $F_\mathrm{bol}$ and $\mu_p$ when estimating sub-Neptune climates and internal structures from observations. 

The super-Earth exoplanet \PiMen\,c orbits a G-dwarf\footnote{$\pi$\,Mensae is also known as HD\,39091.} with an $F_\mathrm{bol} =308\pm7\mathrm{\,So}$ and $T_\mathrm{eff}=5998\pm62\mathrm{\,K}$ alongside two other planets, including a misaligned gas giant \citep{xuan_evidence_2020}. Observations of carbon ions escaping \PiMen\,c and a non-detection of Lyman-$\alpha$ absorption are both suggestive of it hosting a hydrogen-poor atmosphere \citep{Munoz_pimenc_2021, Hatzes_pimenc_2022}. To depart from the low-metallicity end-member of TOI-421\,b, we perform retrievals on \PiMen\,c planet which adopt phenomenological constraints on its composition derived from photochemical-escape models: $\mu_p\ge \SI{4.1}{\amu}$ \citep{Munoz_pimenc_2021}. To incorporate the lower bound on $\mu_p$ within our retrieval, we set $\ln \mathcal{L}=-\infty$ for MCMC samples which violate this $\mu_p$ inequality. We perform four separate retrievals on \PiMen\,c corresponding to each combination of including (or excluding) both $F_\mathrm{bol}$ and $\mu_p$ as observationally informed constraints.

\begin{figure}
    \centering
    \includegraphics[width=\linewidth]{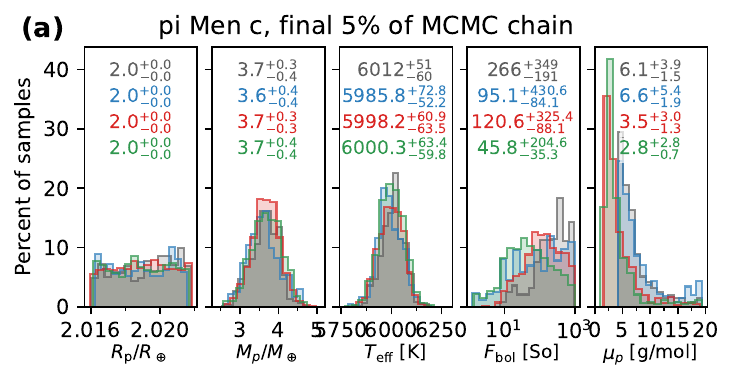}%
    \vspace*{-1mm}
    \includegraphics[width=\linewidth]{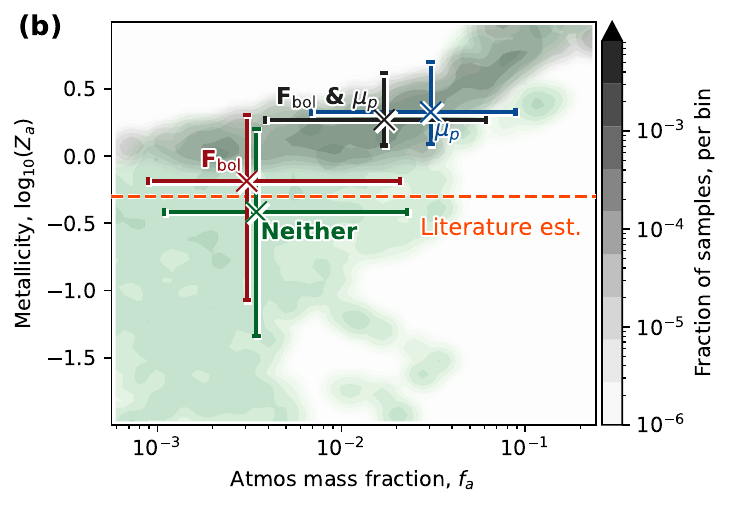}%
    \caption{
    Four separate retrievals (scatter points) on the atmospheric parameters of \PiMen\,c, with and without constraints on atmospheric molecular weight $\mu_p$ and bolometric instellation $F_\mathrm{bol}$.  \textbf{Top}: MCMC posteriors on the constraining observables, for each of the four constraint combinations (colours). \textbf{Bottom}: median values and $\pm 1 \sigma$ ranges on the retrieved  metallicity (y-axis) and atmospheric mass fraction (x-axis), under each combination of constraints (marker colour). The contours in panel (b) visualise the posterior-space of the `Neither' (green) and `$F_\mathrm{bol}$\&$\mu_p$' (black) constraint combinations. The horizontal orange line is the minimum atmosphere molecular weight informed by phenomenological comparisons between photochemical-escape modelling and the detection of C\,II ions escaping from this planet \citep{Munoz_pimenc_2021}.
    }
    \label{fig:ia_Fbol}
\end{figure}

Figure~\ref{fig:ia_Fbol} plots the final 5\% of the MCMC samples from these four retrievals, showing retrieved atmosphere metallicity versus mass fraction in panel (b). The four scatter points show the median and $\pm1\sigma$ ranges on the retrievals' samples. The horizontal orange line is the minimum \ch{H2O} mixing ratio ($Z_a\ge0.5$) informed by photochemical modelling \citep{Munoz_pimenc_2021}. 

We can interpret the structure of \PiMen\,c, using our models, in a functionally equivalent way to generalised mass radius curves by neglecting both $F_\mathrm{bol}$ and $\mu_p$ as constraints on our retrieval (green point, `Neither'). This approach leads to wide estimates on the atmosphere metallicity and mass fraction (panel b). With neither constraint included, the retrieval tends to favour metal-poorer compositions ($\log_{10} Z_a\approx-0.4$) and comparatively less massive atmospheres ($f_a\approx 0.0035$). The flexibility of having left $F_\mathrm{bol}$ as a free parameter makes the solution space highly degenerate, so the forward model can reproduce \PiMen\,c's mass-radius combination for a wide range of atmosphere configurations. Its corresponding posterior distributions in panel (a) explore a wide range of bolometric fluxes and photospheric mean molecular weights (green histograms). 
 
In contrast, introducing $\mu_p$ as an additional observable (blue point in Figure~\ref{fig:ia_Fbol}b) suggests that \PiMen\,c has a substantially more massive envelope, enriched in heavier volatiles. While our derivation of atmospheric mean molecular weights from metallicities is a result of thermochemical calculations, this result remains consistent with the minimum metallicity suggested by photochemical modelling (dashed orange line). However, this case (blue point) still leaves $F_\mathrm{bol}$ as a free parameter and permits large atmosphere mass fractions $f_c$ of up to 8.8\% at $+1\sigma$.  The blue posterior histogram in Figure~\ref{fig:ia_Fbol}a shows the MCMC exploring a wide range of bolometric fluxes when this quantity is not included as a constraining parameter.

\PiMen\,c is exposed to substantial irradiation from its host star, which acts to inflate its radius according to gaseous absorption and scattering processes within its atmosphere. The relevance of these physical effects was demonstrated by the radiative-convective scenarios in our initial case study (solid lines in Figure~\ref{fig:atmosphere}), compared to the simplified $T(p)$ solutions. As a result, when accounting for these physical processes for \PiMen\,c, introducing $F_\mathrm{bol}$ as a constraint on the retrieval shifts our median estimates on the atmosphere mass fraction to lower values (compare red vs green points, or black vs blue points in Figure~\ref{fig:ia_Fbol}). Taking both $F_\mathrm{bol}$ and $\mu_p$ together suggests that \PiMen\,c has an atmosphere comprising approximately 1.7\% of its total mass with a metallicity $\log_{10}Z_a=0.27$ equivalent to $138\times$\,Solar.

Overall, these behaviours emphasise that modelling which self-consistently accounts for a planet's radiation environment can offset systematic behaviours and degeneracies present in related parameters, through their correlated physics. In light of direct indications of ongoing escape from \PiMen\,c's atmosphere, an explanation of its formation scenario and lifetime evolutionary history is directly tied to estimates on its \textit{present-day} atmosphere mass fraction. The inference of a somewhat smaller present-day $f_a$ -- with irradiation environment and radiative transfer physics factored into the retrieval (black point) -- is reflective of a somewhat volatile-poorer formation scenario or less extreme history of escape, given current measurements.

\section{Discussion} 
\label{sec:discuss}

\subsection{Mass-radius relationships arising from climate processes}

We found systematic differences when comparing radii derived from radiative-convective climate calculations to simpler $T(p)$ approximations (Figure~\ref{fig:atmosphere}). Unlike the isothermal scenario, our radiative-convective modelling suggests that super-Earth and sub-Neptune radii are significantly influenced by the local irradiation environment, which alters the atmospheric temperature structure, convective regime, and consequently their photospheric radii. This behaviour is consistent with previous theoretical modelling of larger planets \citep{fortney_planetary_2007, lopez_born_2017, lopez_understanding_2014}. Our suggestion that irradiation, mass, and composition, are of equal importance is supported by empirical studies of exoplanet and the Solar System statistics \citep{ulmer_beyond_2019}. So, while existing planetary static-structure models either neglect the atmosphere's contribution or simplify its temperature structure, our results demonstrate that convectively-stable radiative layers must be considered, because they systematically change photospheric radii compared to purely adiabatic or isothermal scenarios \citep{selsis_cool_2023, peng_puffy_2024, nicholls_convective_2025, cmiel_structure_2025}.  

We suggest that these physically-justified sensitivities to irradiation environment imply that sub-Neptunes cannot be categorised by a single family of mass-radius isolines derived from specific atmosphere compositions or structures (e.g., hydrogen `gas dwarfs' or \ch{H2O} `water world' end-members). Rather, each planet's individual bulk density is a function of its specific stellar context and own composition.  

Further to its influence over planetary radii, our analysis of convective vigour (Section~\ref{sec:results_convect}) shows that a wide range of scenarios may have atmospheres that are largely stable to convection. A spectroscopic treatment of the radiative transfer problem is key to modelling these cases, because the presence of radiative windows between molecular absorption bands -- dependent on chemical composition -- shapes which  scenarios are stable to convection. The abundance of convectively stable scenarios across our parameter space suggests that some exoplanet families could remain compositionally inhomogeneous, even without appealing to de-mixing by chemical immiscibility \citep{howard_the_2025, piaulet_water_2025, vazan_how_2024}.  In the absence of strong convection, higher-order atmosphere dynamics may be necessary to elementally homogenise sub-Neptune atmospheres; e.g., gravity wave breaking \citep{pierrehumbert_book_2010, komacek_vertical_2019}.

Throughout our modelling, we have applied thermochemical equilibrium calculations to calculate vertical speciation from well-mixed elemental abundances. There remains the possibility that, in reality, substantial volatile atmospheres are elementally inhomogeneous. This uncertainty is an important concern because atmospheric processes represent the key connections by which we can infer deep interior processes from the observable upper-atmosphere \citep{katyal_effect_2020, piette_rocky_2023, lichtenberg_constrainin_2025, zilinskas_characteris_2025}. Chemical modelling of sub-Neptunes has previously indicated that upper-atmosphere C/O ratios might not necessarily be representative of their deep interiors \citep{werlen_atmospheric_2025}. In particular, weak mixing may permit chemical stratification of C/O across low-pressure atmospheric regions, rather than quenching the gas composition at near-constant molecular abundances. Chemical kinetics simulations with a fixed weak $K_{zz}=\SI{e4}{\cmsps}$ suggest \citep{werlen_atmospheric_2025} that C/O at the 0.1\,mbar level may be $1000\times$~smaller than the buffered deep C/O value. However, our mixing-length formulation -- which enables a direct estimate of $K_{zz}$ (Figure~\ref{fig:convection}b) -- shows that convective vigour generally corresponds to increased bulk C/O ratios, which could homogenise C/O in most scenarios and effectively marginalise the weak-mixing regime explored by \citet{werlen_atmospheric_2025}. Our results suggest that observed elemental (metallicity) ratios could indeed be reflective of processes happening deep within planetary interiors, enabled by sufficient mixing in the atmosphere and interior-atmosphere equilibration in the deep interior.

It should be noted that we have adopted bulk atmospheric C/O ratio and metallicity $Z_a$ as independent variables on our parameter space, and then calculated atmospheric chemistry profiles self-consistently with their $T(p)$ while setting these two metallicity parameters independently. Redox chemistry within deep planetary interiors -- which become strongly coupled to their atmospheres in the magma ocean regime -- could partially correlate these two parameters \citep{ tian_atmospheric_2024, seo_role_2024,werlen_atmospheric_2025}. A self-consistent incorporation of magma ocean-atmosphere interactions (e.g., volatile degassing) into our retrieval framework would represent another form of parameter-constraining physics. 

Section~\ref{sec:results_gravity} showed that approximating gravitational acceleration as being constant with altitude -- common to spectroscopic retrieval and radiative transfer codes -- is fundamentally limiting. Because $g$ varies non-linearly with altitude, particularly in low-gravity or highly-irradiated regimes, simulations must integrate the mass-continuity and hydrostatic equations self-consistently with their climate modelling \citep{arnscheidt_Atmospher_2019}. 

Stellar X-ray and UV irradiation can efficiently drive hydrodynamic escape by doing work to remove weakly-bound gas parcels \citep{hunten_escape_1987, chassefiere_hydrodynami_1996, yoshida_escape_2024}. This process of `photoevaporation' is thought to explain the small planet radius valley \citep{owen_evap_2017, owen_review_2019, rogers_road_2025}. The minimum flux of X-ray plus ultraviolet radiation required to drive a hydrodynamically escaping outflow depends on upper-atmospheric composition, temperature, and vertical structure \citep{ brain_atmospheric_2016, schulik_aiolos_2023,ji_the_2025}. Importantly, this relationship between the hydrodynamics and structure means that the strength of gravity which  binds a planet's upper-atmosphere plays an important role \citep{ GronoffReview, modirrousta-galian_three_2022,chatterjee2024}. Our results highlight the sensitivity of gravitational attraction $g_p$ to atmosphere mass $f_c$ and a planet's global structure (Section~\ref{sec:results_gravity}). Taken together, these points suggest that population-level predictions for atmosphere retention (or loss) must carefully consider modelling choices of whole-planet structures \citep{rogers_road_2025,ji_the_2025}. Whether a particular planet readily retains (or loses) its atmosphere could be determined by the size of its mantle, in addition to its reservoir of volatiles and exposure to ionising radiation \citep{rogers_redif_2025}.

\subsection{Breaking degeneracies in future exoplanet characterisations}

Mass-radius relations presented in Section~\ref{sec:results_curves} cover a wide range atmosphere compositions, structures, and radiation environments. Traditional static-structure calculations \citep[e.g.][]{seager_mass_2007,Zeng2019, huang_magrathea_2022} either neglect atmospheres entirely, or are founded upon isothermal or adiabatic atmosphere profiles with a narrow set of compositions. Our library of models reproduces the compositional `end-member' cases of  \ch{H2}- or \ch{H2O}-dominated atmospheres. For instance, high-altitude scale heights for \ch{H2}-rich atmospheres on Earth-like interiors match theoretical expectations for inflated gas-dwarfs \citep{rogers_road_2025}. However, our models also explore diverse compositions to make the emerging metal-rich exoplanet regime accessible to structure modelling \cite[Figure~\ref{fig:chem};][]{kama_abundant_2019, Suer2023, sossi_review_2025}. The masses, thermodynamics, and radiative properties of these molecules shape the climate states of exoplanetary atmospheres, and thus their observable radii (Figures~\ref{fig:atmosphere}~and~\ref{fig:convection}).

Observational estimates on planetary metallic-core sizes are valuable; the presence and size of a planet's core is key to understanding a planet's historical and present-day mantle redox state, because the pressure-temperature conditions at the core-mantle boundary determine chemical equilibration between volatile elements and metallic phases \citep{wade_core_2005, lichtenberg_redox_2021, guimond_mineralogical_2023, young_diff_2025}. Planet formation modelling and empirical trends from the Solar System suggest a gradient of decreasing planet core size and increasing mantle oxidation for greater planet-star separations \citep{sossi_review_2025}, however, estimation of planet core sizes remains difficult -- even for our Solar System neighbours \citep{khan_mars_2018, jacobson_formation_2017, smrekar_venus_2018}. Recent modelling has suggested that planets with interior masses $M_s>4M_\oplus$ may lack distinct metallic cores because their iron is largely oxidised into \ch{FeO}, which is readily incorporated into mantle-building material \citep{huang_limits_2025}.  Regardless, even first-order bounds on the structure of deep exoplanet interiors would provide valuable insight on their formation, migration, and redox state -- potentially indicative of evolutionary pathways different from Earth's \citep{lichtenberg_redox_2021}. While our demonstrations with \PiMen\,c and TOI-421\,b do not strongly constrain core fractions $f_c$, the mass-radius isolines presented in Figure~\ref{fig:mr_c} suggest that -- when accounting for atmosphere inflation by its irradiation environment -- the radii of lower-mass exoplanets could remain indicative of $f_c$. 

Observations of specific planets interpreted in previous studies have applied static-structure calculations which do account for particular planetary irradiation environments \citep{piaulet_steam_2024}. However, it is remains common practice within the literature to directly compare a planet's mass and radius with generalised isolines on a two-dimensional parameter space;  \citep[e.g.,][as recent examples]{meech_jwst_2026, wallack_jwst_2026, palethorpe_constrainin_2026}. To make these two-dimensional plots, planetary structure models must be necessarily insensitive to -- or at least projected from -- a given temperature or irradiation environment.  We emphasise that caution must be exercised when placing specific planets on this two-dimensional space, even when comparing against isolines which correspond to a spectrum of atmospheric compositions. 

For some parameters, such as $\mu_p$, there are multiple physical processes and planetary characteristics contriving to set the values which we measure. It is therefore easier to place observational constraints on some parameters compared to others (e.g., $\mu_p$ compared to the mixing ratios of specific species). As a result, `free' retrieval techniques have become commonplace \citep{welbanks_aurora_2021, barstow_outstanding_2020}. These techniques permit physically-correlated parameters to be varied independently of each other. 

For example, a planet's photospheric gravity might be treated independently of the temperature structure and composition of the whole atmosphere. However, in reality, the physics might dictate that an apparently-low photospheric gravity is only physically compatible with an \ch{H2}-rich composition, enabled by large scale heights and photospheric radii. The physical correlations between atmosphere mass fraction, metallicity, and planetary instellation (Figure~\ref{fig:mr}) caution against free retrieval approaches. Indeed, the application of physically unconstrained retrievals to the case of K2-18\,b has been criticised in the literature \citep{welbanks_challenges_2026, schmidt_comprehensive_2025}, since  relaxing physical constraints \citep[e.g., on albedo and composition;][]{barrier_general_2025} permits biosignature claims for this planet. Instead, comprehensive chemistry-climate models can be readily integrated into fast static-structure retrieval frameworks (Section~\ref{sec:results_ia}) and thereby allow richer insight deep into planetary interiors and their evolutionary histories.  We can leverage these models to dampen systematic trends between otherwise `free' parameters, rather than view these physical effects as computational limitations or as known-unknown factors \citep{welbanks_challenges_2026, hammond_reliable_2025}.

In particular, our demonstrations with TOI-421\,b and \PiMen\,c show that including irradiation and molecular weight as constraints can improve estimates of planetary properties compared to those returned from free retrieval or from direct comparisons in mass-radius space. In the case of \PiMen\,c, the introduction of bolometric instellation flux $F_\mathrm{bol}$ as a constraint suggests a relatively less massive but more metal-rich atmosphere than a direct comparison with mass-radius curves would predict \citep{Zeng2019}. 

In addition to climate and irradiation physics, another opportunity to leverage physical sensitivities is through the non-linear relationship between photospheric gravity (or scale height) and atmosphere mass fraction, which is shown in Figure~\ref{fig:gravity}. A pathological example of this effect would be the observation of shallow transmission feature depths (suggestive of high metallicity) on a low-density planet (suggestive of low metallicity). In this case, the gravitational attraction of a substantially massive atmosphere could explain the muted spectroscopy, despite a hydrogen-rich atmospheric composition. Accounting for these physical interactions requires the incorporation of vertically resolved forward-models of whole-planet structure into retrieval frameworks (Section~\ref{sec:methods_infer}). 

\subsection{Surface environments of sub-Neptune worlds}

The sub-Neptune exoplanet population is upper-bounded by Neptune's radius  ($<3.86R_\oplus$) and lower-bounded by the `radius valley' \citep{lopez_understanding_2014, fulton_california-_2018, rogers_most_2015}. The radius valley is a weak function of orbital period and stellar type \citep{ho_shallower_2024, Venturini2024, gillis_tess_2026}, so rigorously categorising a planet as being a sub-Neptune is difficult. Importantly, the radius valley is thought to arise from the interaction of planetary formation scenarios (interior or exterior to the water ice line), escape processes (boil off, core powered mass loss, photoevaporation), and planetary migration \citep{rogers_road_2025, Venturini2020, Burn2024}. These processes link sub-Neptunes to the super-Earths below the radius valley \citep{nicholls_escape_2025, tang_cpml_2024, owen_evap_2017}; radius-age trends from the Kepler-California survey suggest that, since sub-Neptunes are generally younger, they may represent progenitors of the super-Earth population \citep{david_cks_2021, owen_evap_2017}.

What are sub-Neptunes? We may expect the small exoplanet population to comprise multiple subfamilies of diverse provenance, arising from a continuum of formation scenarios and lifetime histories \citep{lichtenberg_constrainin_2025, kite_atmosphere_2020, bean_nature_2021}. These may include \ch{H2O}-dominated or \ch{H2}-dominated end-members, as well as the emerging population of metal-rich worlds \citep{belloarufe_evidence_2025, benneke_jwst_2024, nicholls_escape_2025, Kempton2023}. Independent of this diversity, low bulk densities can only be explained by substantial inventory of volatiles ($f_a$) and/or planets inflated by stellar irradiation ($F_\mathrm{bol}$). Whether by large $f_a$ or $F_\mathrm{bol}$, the physics of atmospheric energy transport then suggests that sub-Neptune exoplanets should readily host hot surfaces and interiors.

Figure~\ref{fig:hz} empirically probes the range of simulated surface conditions arising from our grid of models, showing smaller planets in the top panel (a) and sub-Neptunes in the bottom panel (b). Regardless of their radii, the majority of our cases proximate to the canonical habitable zone exhibit uninhabitable surface temperatures which exceed the critical point of water \citep{innes_runaway_2023, habib_convection_2024}. This result is consistent with modelling of TOI-270\,d \citep{benneke_jwst_2024} and K2-18\,b \citep{shorttle_k218b_2024, luu_can_2024}. Sub-Neptunes near traditional habitable zone boundaries (panel b) are unlikely to host clement surface conditions because of their substantial optically-thick atmospheres, potentially undermining the proposed `hycean' scenario \citep{madhusudhan_chemical_2023}. Any additional internal heat production, such as from tidal or induction heating, would further increase their surface temperatures beyond our conservative estimates \citep{nicholls_tidal_2025, Kislyakova_induction_2023, herath_thermal_2024, van_onset_2025}. A window for surface habitability on sub-Neptunes may exist at larger orbital separations \citep{pierrehumbert_hydrogen_2011} or within cooler cloudy environments far above their surfaces \citep{delort_clouds_2017, limaye_clouds_2018}. As with direct mass-radius isolines, comparisons between specific sub-Neptune exoplanets and habitable zone boundaries must be undertaken carefully. Unification of planet-structure modelling with the habitable zone framework, via $R_p$ in Figure~\ref{fig:hz}, shows that rapid interpretive frameworks can remain useful given a treatment of the climate physics applicable to the sub-Neptune regime.

Figure~\ref{fig:hz}b suggests that sub-Neptunes may have largely molten interior magma oceans \citep{calder_most_2025}, which enables the efficient exchange of energy and material between their interiors and atmospheres \citep{elkins_linked_2008, hirschmann_magma_2012, schaefer_review_2018}. Since the large radii of sub-Neptunes make them especially amenable to characterisation with current observatories, these planets represent an opportunity for probing the geochemistry and geodynamics deep within exoplanet interiors \citep{dorn_structure_2015,  bean_nature_2021,lichtenberg_constrainin_2025}. Hydrogen thermochemistry generally favours the formation of either \ch{H2} or \ch{H2O}, depending on the availability of oxygen, thereby tracing metal-poor and metal-rich formation scenarios \citep{krijt_chemical_2023}. However, sulfur chemistry and the unique interior-atmosphere partitioning behaviour of S represents a parallel opportunity for probing planetary histories and interior environments. $\mathrm{S}^{2-}$ is removed from atmospheres by dissolution into silicate melts at reducing conditions \citep{namur_mercury_2016, boulliung_so2_2022}, so non-detection of sulfur-bearing molecules within sub-Neptunes atmospheres could be taken as evidence for S storage in a magma ocean \citep{shorttle_k218b_2024}.  

Although we have conservatively adopted a solar S/H ratio, sulfur-bearing species \ch{H2S} and \ch{SO2} are found to be abundant across our chemical equilibrium calculations (Figure \ref{fig:chem}). The outsized influence of sulfur species over total atmospheric mean molecular weight -- through the comparatively large mass of S atoms -- and their chemical abundance highlights their observational accessibility, given models capable of sufficiently treating the relevant chemistry and correlated physics.  Our whole-planet retrievals incorporate simultaneous constraints on bulk structure, irradiation, and composition; these are collectively necessary to distinguish between degenerate physical scenarios.

In particular, our demonstration of whole-planet structure retrieval with TOI-421\,b suggested that its atmosphere gases are fully-miscible above a molten interior (Section~\ref{sec:results_ia_demo}). This result is consistent with a non-detection of \ch{NH3} and the suggestion of \ch{SO2} in its atmosphere from JWST spectroscopy \citep{davenport_toi421b_2025}. Similarly, the super-Earth L\,98-59\,d has a reducing atmosphere with suggestions of photochemically-produced \ch{SO2} \citep{nicholls_escape_2025}. Evolutionary models should test whether these two under-dense planets have experienced similar life-time evolution pathways, and whether they are members of a shared population of reduced magma ocean worlds.

\subsection{Future targets for model development}

Our library of models is an advancement over the existing set of openly-available libraries of static structure which relate exoplanet compositions, masses, and radii. However, no model is ever `complete' or entirely representative of reality \citep{box_science_1976}. Several physical complexities remain open for future exploration. 

A primary consideration is the role of aerosols -- such as clouds -- which could significantly alter the Bond albedo of these planets (impacting their climatic energy balance and temperature structure) and mute molecular features probed by transmission spectroscopy \citep{barstow_outstanding_2020, Kempton2023}. Neglecting these effects will potentially lead to overestimates of atmospheric scale heights \citep{Morley2015, gao_aerosol_2020}. Similarly, our current chemical model focuses on the thermochemical equilibrium regime, however, the interplay between photochemistry and vertical mixing processes could drive the upper atmosphere out of equilibrium \citep{madhusudhan_exoplanetary_2016, tsai_comparative_2021, venot_chemical_2012, nicholls_temperaturechemistry_2023} -- particularly for  planets exposed to large UV fluxes, and given our indications of weak vertical-mixing (Section~\ref{sec:results_convect}). 

Regarding the planetary interior: our modelling considers distinct boundaries between the atmosphere, mantle, and core. Informed by the Solar System planets, this approach has been widely adopted in planetary models. For higher mass planets, these interfaces may be better represented by diffuse transitional regions due to the potential for atmosphere-silicate or silicate-core miscibility \citep{stevenson_interiors_1982, Vazan2018, benneke_jwst_2024, schlichting_chemical_2022, gilmore_coreenvelo_2026}. Equilibration between core, mantle, and atmosphere in $M_p>6M_\oplus$ sub-Neptune planets is supported by evidence from \textit{ab-initio} molecular dynamics simulations, for simplified ternary mixtures \citep{young_diff_2025, gupta_the_2025}. Miscibility between these domains would permit substantial incorporation of hydrogen into planetary cores, resulting in simultaneous decreases in core density and planet radius \citep{young_diff_2025}. The resultant sensitivity of planet radius $R_p$ demonstrated by these initial studies on structural miscibility suggests that these effects are of equal importance to the sensitivities in atmospheric structure demonstrated  by Sections~\ref{sec:results_atmosphere}~and~\ref{sec:results_curves}.

Finally, expanding our grid to include more exotic interior architectures would further refine our ability to break observational degeneracies in the sub-Neptune population. We might consider carbon-rich mantles, broader core-to-mantle mass ratios $f_{ci}$, or the presence of high-pressure water-ice layers \citep{Sotin2007, Madhusudhan2012, Burn2024, chakrabarty_waterworld_2024, valencia_diversity_2025}. These factors represent suggestions for future development which builds upon our current results, as they could be directly incorporated into future iterations of these calculations.

\subsection{Evolutionary retrievals to supersede static-structure modelling}

We have shown that planet- or system-specific parameters are central in connecting theoretical planetary structures and compositions with masses and radii. However, even with fast forward-models wrapped around large pre-computed libraries, an increasing number of parameters will inevitably add computational expense \citep{fisher_how_2022, madhusudhan_bayes_2009}. Furthermore, static models are directly sensitive to hystereses arising from the physics of planetary evolution \citep{lichtenberg_chili_2025}. For example, \ch{H2}-rich atmospheres are unlikely to survive around small interiors for billions of years due to escape processes which readily strip low-metallicity envelopes \citep{lopez_born_2017, ji_the_2025, zahnle_cosmic_2017, modirrousta-galian_three_2022}. A self-consistent incorporation of escape modelling into planetary retrieval frameworks would naturally disfavour a subset of parameter-space, providing tighter constraints on atmosphere mass fractions, metallicities, and the corresponding surface environments \citep{krissansen_erosion_2024, luger_extreme_2015}. The case of TOI-270\,b represents a valuable application of such models \citep{benneke_jwst_2024}. Since static models cannot incorporate these physical processes, they will not universally respect the bounds on the `true' set of scenarios accessible to a planet during its lifetime \citep{lichtenberg_chili_2025, aguichine_evolution_2024, nicholls_escape_2025}.

Accurate and fast models are required for interpreting observations. Ultimately, the future may lie in numerical models of planetary \textit{evolution} which self-consistently represent the physics which shapes planets throughout their complete lifetimes. For example, our demonstrations of a retrieval on super-Earth \PiMen\,c are consistent with a wide range of present-day atmosphere mass fractions, but application of evolutionary modelling could tightly bracket the set scenarios which this planet could have reasonably attained. Time-resolved simulations of planetary evolution also naturally treat the `problem' of internal heat production because they calculate this quantity as a dependent output variable, rather than leaving it a free parameter requiring estimation.

A limiting characteristic of evolutionary models is their computational expense. As part of the wider CUISINES intercomparison framework \citep{sohl_cuisines_2024}, the CHILI project provides a quantitive comparison between static climate calculations and fully-coupled evolutionary simulations of planetary histories \citep{lichtenberg_chili_2025}. These direct comparisons between static `snapshots' and evolutionary calculations will inform how best to approach the future of planetary retrievals. Historically, evolutionary models have only been applied for theoretical case studies or to specific planets through expensive grid searches. However, in the case where CHILI finds that evolutionary simulations are necessary for accurate modelling of past and present (exo)planetary environments, we suggest that evolutionary codes \citep[e.g., the PROTEUS framework;][]{lichtenberg_vertically_2021,nicholls_redox_2024} could be liberally applied through retrieval techniques.  This future could take the form of large model libraries being used to train machine learning emulators, or evolutionary models applied via traditional inference methods.

\section{Conclusion} 
\label{sec:conclude}

In this work, we have moved beyond simplified mass-radius relations to provide an assessment of whole-planet static structure, through self-consistent radiative-convective-chemical equilibrium modelling. We constructed a grid of 504,000 different scenarios which could feasibly arise from the combined processes of planet formation and Gyr-scale evolution.

Our findings lead to the following key conclusions:

\begin{enumerate}
    \item Comprehensive climate calculations are a necessary component of planetary structure models. We have shown that radiative-convective equilibrium solutions, which account for radiative windows and the role of stellar irradiation, are essential for accurate characterisation from bulk planet properties because observed radii can be sensitive to instellation flux in addition to variations in core-envelope proportions. Traditional isothermal and adiabatic climate approximations can systematically misestimate planetary radii on a population-level.

    \item Gravity is not a simple function of radius. We have demonstrated that adopting a constant-gravity approximation is limiting, particularly when modelling atmospheres that have large scale heights or masses. The non-linear variation of gravity with altitude can significantly influence upper-atmosphere scale height and, by extension, the perceived bulk density and transit feature depths.

    \item A substantial proportion of small exoplanets -- even those within the traditional habitable zone -- may host surface conditions exceeding the critical point of water. This theoretically narrows the set of regimes in which exoplanets can support  surface water oceans. Instead, a magma ocean regime may be common place, which offers multiple opportunities for using atmospheric characterisations to probe deep interiors via interior-atmosphere interactions. Habitable regimes may exist at low instellations beyond the habitable zone, highlighting the need to look beyond the instellation-$T_\mathrm{eff}$ axis for surface characterisation, and apply climate models appropriately to specific scenarios.

    \item To make our modelling accessible, we have developed the `InferAGNI' Python package: a Bayesian inference tool which leverages our extensive library of planet structure models via an MCMC sampling algorithm. Our demonstrations with TOI-421,b and \PiMen\,c illustrate that incorporating physical constraints on atmospheric structure and instellation can lift degeneracies inherent in direct mass-radius interpretations.

\end{enumerate}

The scientific community is preparing for the upcoming wealth of data to be delivered by PLATO, Ariel, and the ELT. Modelling is necessary for interpreting these measurements, so a transition from simplified static-structure calculations to consistent climate-structure modelling is required. JWST has already characterised exoplanetary environments inaccessible  to existing planet-structure models. Our library of simulations and open-source tooling provide an expanded framework for interpreting these observations. The future of exoplanet science may necessitate wider application planetary evolution models in interpreting observations, rather than simulating static snapshots. Evolutionary simulations can resolve the history of atmospheric escape and interior degassing, ultimately explaining the diverse origins of exoplanets in our galaxy.

\section*{Acknowledgements}

H.N. and O.S. acknowledge support from STFC grant UKRI1184. H.N. thanks Raymond T. Pierrehumbert for guidance during the initial stages of this project, and also thanks Robb Calder for his valuable feedback on the paper draft. T.L. was supported by the Branco Weiss Foundation, the Alfred P. Sloan Foundation (AEThER project, G202114194), NASA's Nexus for Exoplanet System Science research coordination network (Alien Earths project, 80NSSC21K0593), and the European Research Council (ERC) under the European Union's Horizon Europe research and innovation programme (101219807, MagmaWorlds). This research has made use of data obtained the portal \url{exoplanet.eu} of The Extrasolar Planets Encyclopaedia. We thank the Center for Information Technology of the University of Groningen for their support and for providing access to the H\'abr\'ok high performance computing cluster. We thank our reviewer for their constructive suggestions and thoughtful feedback.

% https://www.elsevier.com/researcher/author/policies-and-guidelines/credit-author-statement
CRediT author statements.
\textbf{HN}:  Conceptualization, Methodology, Software, Formal analysis, Investigation, Data Curation, Writing - Original Draft, Visualization.
\textbf{TL}:  Conceptualization, Methodology, Software, Writing - review \& editing.
\textbf{OS}:  Conceptualization, Writing - Review \& Editing, Supervision, Funding acquisition, Project administration.
\textbf{FP}:  Methodology, Software.

%%%%%%%%%%%%%%%%%%%%%%%%%%%%%%%%%%%%%%%%%%%%%%%%%%
\section*{Data Availability}
\label{sec:dataavail}

The data underlying this article are available in Zenodo, at \url{https://dx.doi.org/10.5281/zenodo.19193681} \citep{inferagni_zenodo}. Our retrieval tool source code is available on GitHub under the free and open source GPLv3 licence, at \url{https://github.com/nichollsh/InferAGNI}. InferAGNI is also accessible via the Python Package Index (PyPI), at \url{https://pypi.org/project/InferAGNI/}.


\bibliographystyle{mnras}
\bibliography{main.bib}

%%%%%%%%%%%%%%%%% APPENDICES %%%%%%%%%%%%%%%%%%%%%

\appendix

\section{Exoplanet statistics}
\label{app:exoplanets}

For the purposes of informing the case study in Section~\ref{sec:results_atmosphere} and observational analysis in Appendix~\ref{app:transspec}, and for contextualising the remainder of our Results and Discussion, this appendix presents the statistics of the surveyed small exoplanet population. These data were sourced from \url{exoplanet.eu} on 27\,February\,2026, which tabulates 6260 exoplanet candidates. Since this present study is specifically concerned with the structure of `small' exoplanets, we only consider planet candidates which fulfil the criteria below:
\begin{itemize}
    \item  labelled as `confirmed' in the \url{exoplanet.eu} database,
    \item have quantified radii (i.e., planets which have been measured via transit methods),
    \item have mass estimates less than Neptune's mass (i.e., $17M_\oplus$ -- \citet{lodders_planetary_1998}).
\end{itemize}
These filters together yield a sub-population of 753 `small' exoplanets. 

\begin{figure}
    \centering
    \includegraphics[width=\linewidth]{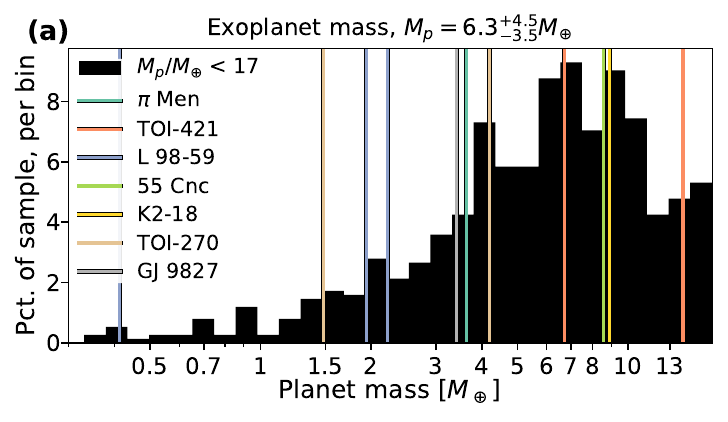}%
    \vspace*{-1mm}
    \includegraphics[width=\linewidth]{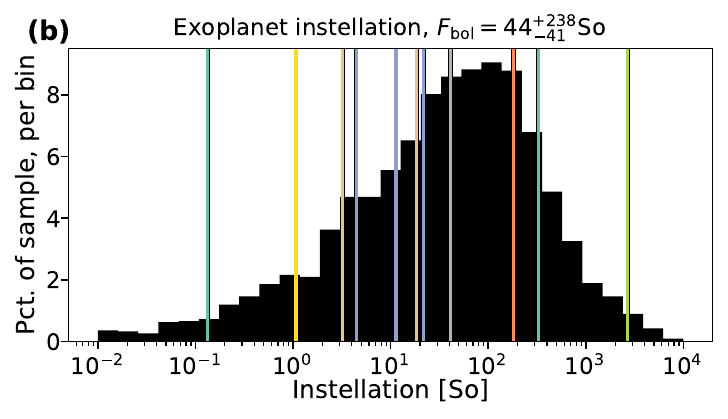}%
    \vspace*{-1mm}
    \includegraphics[width=\linewidth]{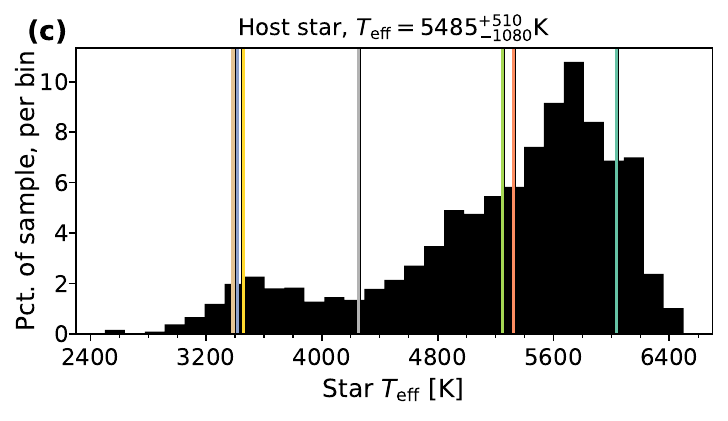}%
     \vspace*{-1mm}
    \includegraphics[width=\linewidth]{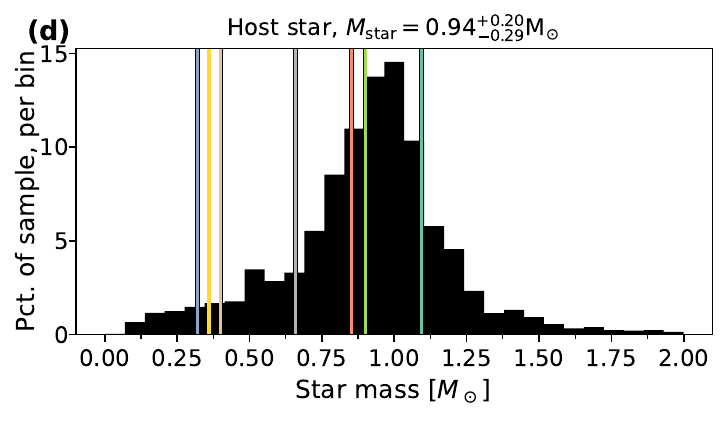}%
    \caption{Histograms of planet mass (\textbf{a}), instellation (\textbf{b}), host star effective temperature (\textbf{c}), and host star mass (\textbf{d}) for confirmed exoplanets with masses less than Neptune ($17 M_\oplus$). Data sourced from \url{exoplanet.eu} on 27\,February\,2026. The overplotted vertical lines indicate values exoplanets of particular interest.}
    \label{fig:exoplanets}
\end{figure}

Figure~\ref{fig:exoplanets} plots the distributions of total planet mass (panel \textbf{a}), bolometric instellation flux (panel \textbf{b}), host star effective temperature (panel \textbf{c}), and host star mass (panel \textbf{d}) for surveyed exoplanets which satisfy the criteria above.  The median values and $\pm1\sigma$ standard deviations on each variable are quantified above each panel of the figure.  Vertical coloured lines (overplotted) highlight exoplanet systems of particular interest for this study. 

It is useful to define a representative `median' small exoplanet for the purposes of case studies in Section~\ref{sec:results_atmosphere} and Appendix ~\ref{app:transspec}. To do this, we adopt the median values  on the distributions plotted in Figure~\ref{fig:exoplanets}.  This `median' exoplanet has a total mass of $6.3M_\oplus$, instellation $F_\mathrm{bol}=44$\,So, and orbits a host star with effective temperature $T_\mathrm{eff}=5486$\,K and mass $0.94M_\odot$. This median exoplanet is defined for the purposes of performing our case studies, although different approaches could be taken to derive a representative test exoplanet. This configuration could be interpreted as a warm sub-Neptune orbiting a G7-type star \citep{kippenhahn_stellar_2012, baraffe_new_2015}.

\section{Equilibrium chemistry}
\label{app:chem}

Atmospheric composition directly sets its opacity to radiation, which we can measure using remote sensing methods; e.g., transmission spectroscopy \citep{barstow_comparison_2020, madhusudhan_atmospheric_2018}. Spectroscopically-varying gas opacities determine when and where absorption of stellar radiation triggers atmospheric convection \citep{pierrehumbert_book_2010, cmiel_structure_2025}. In this work, we perform self-consistent calculations of radiative-convective climates alongside equilibrium thermochemistry. For a given set of elemental abundances (Section~\ref{sec:methods_planet}), the chemical composition at each layer of the atmosphere is calculated using FastChem at to the local temperature and pressure. Feedback between chemistry and climate is known to be important for shaping atmospheric structure \citep{venot_chemical_2012, stock_faschem_2022, nicholls_temperaturechemistry_2023}. 

\begin{figure}
    \centering
    \includegraphics[width=0.98\linewidth]{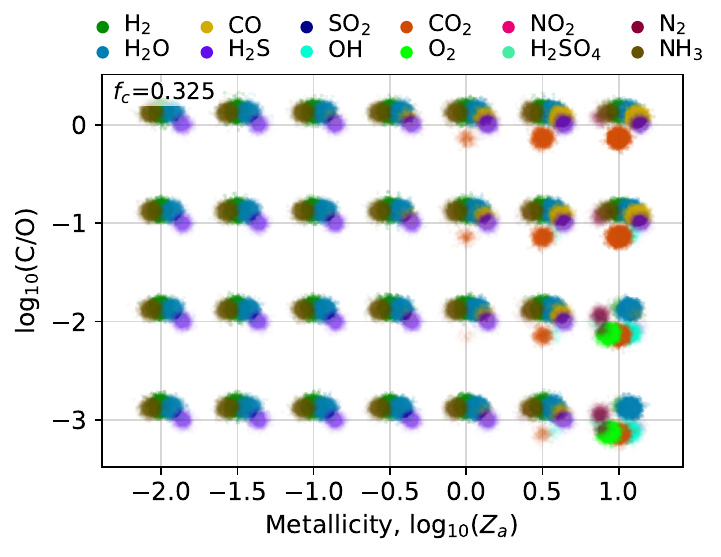}%
    \caption{
    Qualitative representation of the chemical mixtures (colours) which arise across our compositional parameter space, as functions of C/O~and $Z_a$ (axes). These correspond to chemical mixtures at the bottom-most atmosphere layer (i.e., of gases outgassed at the surface). The logarithmic volume mixing ratio of each gas (marker colours) is denoted by their opacity in the plot. 
    }
    \label{fig:chem}
\end{figure}

Figure~\ref{fig:chem} qualitatively visualises the chemical compositions of our model atmospheres. For each C/O--$Z_a$ combination, a scatter point is plotted for each gas which has a volume mixing ratio $\chi$ exceeding $10^{-4}$. The opacity of each point scales logarithmically between $\chi=10^{-4}$ and $\chi=1$.

The low-metallicity regime is dominated by hydrogen (green blobs), with additional contributions from water (blue), ammonia (brown), and hydrogen sulfide (purple). The compositions become increasingly diverse at higher metallicities: \ch{CO2}, \ch{CO}, and \ch{O2} become increasingly abundant. 

\begin{figure}
    \centering
    \includegraphics[width=0.98\linewidth]{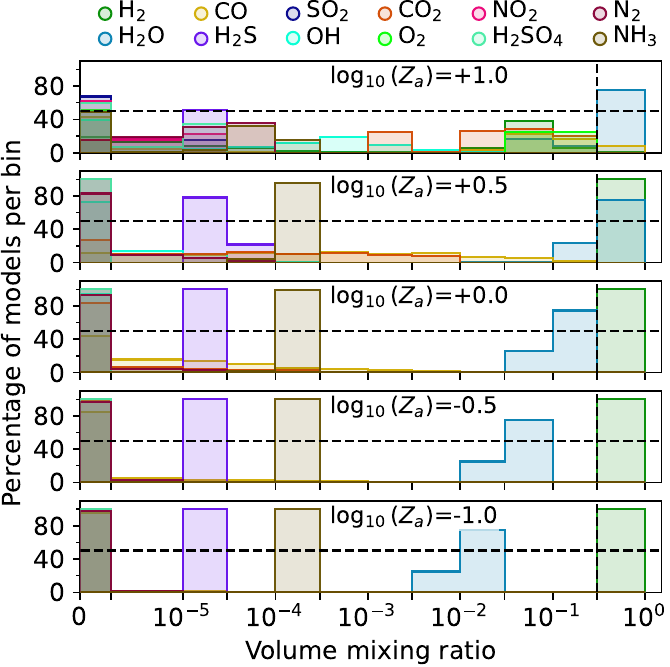}%
    \caption{
    Histograms quantifying the relative dominance of chemical species (colours) across our range of metallicity regimes (panels). For each metallicity $Z_a$, each histogram plots the distribution of volume mixing ratios that arise for a given species (colours). Dashed black lines divide the space into four quadrants, divided between 50\% of the modelled outcomes (y-axis), and dominant versus sub-dominant species (according to logarithmic mixing ratios, $\chi=-\log_{10}(50\%)=0.30$, x-axis). 
    }
    \label{fig:chem_hist}
\end{figure}

Further to our qualitative analysis of compositional diversity, we quantify the statistical trends in thermochemical speciation that arise across our grid, as a function of the total atmospheric metallicity $Z_a$. Each panel of Figure~\ref{fig:chem_hist} slices our grid through a different metallicity regime, for which the coloured histograms plot the statistical diversity in species' outgassed volume mixing ratio. 

We marginalise over C/O ratio, so all of these cross-sections through our metallicity axis are dominated by either \ch{H2} or \ch{H2O}. Atmospheric compositions with $\log_{10}(Z_a)=0.5$ are a largely a binary mixture of dominated by \ch{H2} and \ch{H2O}, indicated by the green and blue histograms entering the upper-right quadrant of the second panel. Lower-metallicity regimes becoming increasingly \ch{H2}-dominated (lower panels).

The highest metallicity cross-section (top panel of Figure~\ref{fig:chem_hist}) shows substantial diversity, reflecting the qualitative variance shown in the right-most column of Figure~\ref{fig:chem}. While not dominant, the abundances of \ch{CO}, \ch{O2}, and \ch{CO2} (yellow, lime, orange histograms) are substantial; comprising >1\% of these metal-rich atmospheres' compositions. 

We model the thermochemistry of nitrogen- and sulfur-bearing species while adopting the Solar photospheric composition as a conservative lower-limit on atmospheric abundances of N and S atoms (Section~\ref{sec:methods_compose}). The oxidised forms of these elements arise in the metal-rich regime (top panel of Figure~\ref{fig:chem_hist}) as \ch{SO2} and \ch{NO2} with minor -- although finite -- abundances of less than 0.01\%. So, although these oxidised molecules have strong infrared absorption cross-sections, they have negligible influence on the planetary radii and structures presented in Section~\ref{sec:results} \citep{boulliung_so2_2022, tennyson_exomol_2018}. 

The abundances of \ch{H2S} and \ch{NH3} in our modelled atmospheres do not depend strongly on the total metallicity (purple and brown histograms) because these reduced forms of S and N also bear H atoms, and so remain stoichiometrically favourable at small $Z_a$. Their modelled volume mixing ratios remain less than 0.1\%. However, given that our adopted S/H and N/H ratios represent reasonable lower limits, future modelling could explore how variance in S/H and N/H promote diversity in the thermochemical formation of \ch{H2S} and \ch{NH3}.

\section{Planet radii probed with transit spectroscopy}
\label{app:transspec}
The radius of an atmosphere-hosting exoplanet is not a well-defined quantity \citep{dotson_exoplanets_2011}. Firstly, because atmospheres are diffuse regions without a distinct upper-boundary between gas and `Space'. Secondly, because exoplanets are not spatially resolved by current instruments. Thirdly, because planets are not spherically-symmetric or one-dimensional \citep{van_retrieving_2025, grant_trans_2023}. Yet, transmission spectroscopy measures the geometric sizes (i.e., the radii) of exoplanets for comparison with modelling, so a definition must be adopted. Transit measurements probe the region of an exoplanet's  atmosphere where stellar irradiation is strongly attenuated by gaseous absorption \citep{madhusudhan_atmospheric_2018}. This is referred to as the `limb' region of an exoplanet, and corresponds to the day-night terminator in tidally locked regimes. The radial location of this photosphere layer is a function of wavelength (since gases absorb spectroscopically), atmospheric temperature structure, and atmospheric composition. 

All current planet structure models are 1D calculations  which aim to be representative of the average structure of a planet \citep{fortney_hot_2021, marley_review_2015, valencia_radius_2007}. It has been common to intuitively define the radial location of the photosphere by a characteristic pressure \citep{tang_cpml_2024, lopez_understanding_2014, huang_magrathea_2022}. For example, hot Jupiter models have considered 1~mbar \citep{fortney_transit_2005} and widely-adopted structure models of \citet{Zeng2019} also used 1~mbar, although  the photosphere region is expected to vary in location from $\sim1$~mbar~to~$~\sim100$s~mbar \citep{aguichine_evolution_2024, huang_magrathea_2022}. 

In our main results, we choose a representative 20~mbar pressure to define the photosphere region (Section~\ref{sec:methods_planet}), following \citet{lopez_understanding_2014}.  This is a simplification which avoids dealing with the wavelength- and composition-dependent factors that determine the photosphere. Self-consistently deriving planet radii based on the optical depths calculated in our climate simulations would not necessarily provide a more robust metric for estimating transit radii from whole-planet structures, because our optical depths correspond to planet-average structure profiles, rather than limb profiles \citep{hubbard_transit_2001, burrows_transit_2003}. Consider  beams of collimated radiation passing through the upper atmosphere following either a vertical (radial) or horizontal (slant) path. \citet{fortney_transit_2005} estimated that the optical depth of specific region,  measured relative to the top-of-atmosphere, will have a slant-path optical depth  35 to 90 times, the radial optical depth depending on composition and temperature. The mass-path is larger for the slant beam than the radial beam. Transmission spectroscopy at wavelength $\lambda$ probes the region where slant optical depth $\tau_s(\lambda)=1$, so the photosphere will exist near $\tau_r(\lambda)=0.01$~to~0.03.  

Previous studies have not robustly incorporated radiative-convective-chemical calculations into interior-atmosphere structure modelling, so did not directly deal with limb versus planet-average differences in structure. A reasonable determination of planet radii remains  a central  consideration this work, so here we extend our main analysis  by connecting representative planet-averaged structures to the conditions probed by transmission measurements.  We perform a set of climate calculations at various metallicities $Z_a$ (following, and determine the region  where $\tau_r=0.02$. The photosphere is estimated for two wavelengths, $\lambda=$\SI{0.675}{\micro\meter}~and~\SI{15.000}{\micro\meter}, corresponding to potential the PLATO~red/blue~filter crossover and JWST-MIRI~F1500W photometric measurements \citep{rauer_plato_2025, alatalo_jwst_2016}.

\begin{figure}
    \centering
    \includegraphics[width=0.98\linewidth]{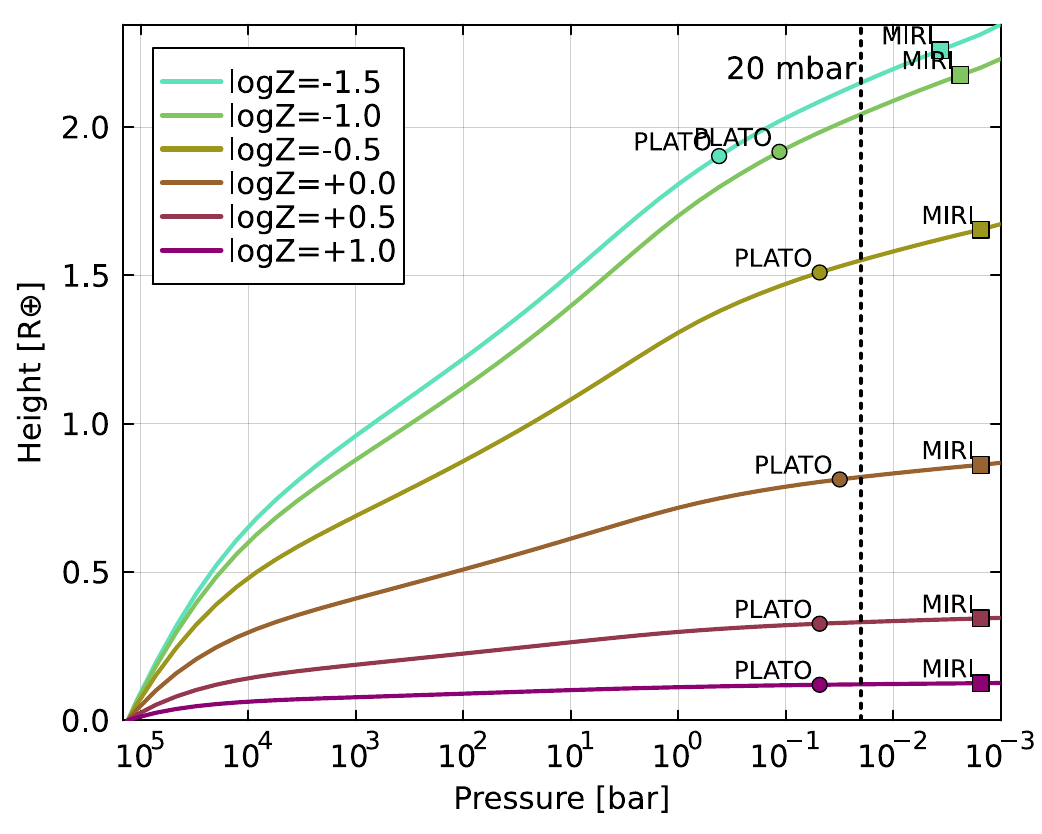}%
    \caption{
    Atmospheric height-pressure structures, modelled for the `median' small exoplanet (Appendix~\ref{app:exoplanets}) under various metallicities (Table~\ref{tab:params}), with two potential locations of its photosphere estimated for each case. Line colours denote atmosphere metallicity $\log_{10} Z_a$ for a binary \ch{H2}+\ch{H2O} composition. The photosphere is estimated for radial optical depths $\tau_r=0.02$ at wavelengths $\lambda=\{0.675, 15.000\}\mathrm{\,\mu m}$ (circular and square points). The dotted black line shows the  20\,mbar layer corresponding to our nominal pressure-defined photosphere (Section~\ref{sec:methods_planet}), which sits comfortably between the PLATO and MIRI photospheres.
    }
    \label{fig:transspec}
\end{figure}

Figure~\ref{fig:transspec} plots atmosphere height-pressure structures and estimated photosphere locations, for these six representative test scenarios (line colours). Our nominal choice of a 20\,mbar photosphere pressure is indicated by the dotted black line. Based on the optical depths for wavelengths probed by PLATO and MIRI photometry (scatter points), it is apparent that our choice of 20\,mbar to define the photosphere reasonably represents a `typical' photospheric region. This result follows from previous literature \citep{lopez_understanding_2014, fortney_transit_2005, dotson_exoplanets_2011}. 

Importantly, the gradient of Figure~\ref{fig:transspec} height-pressure profiles is small: pressure variations across multiple orders of magnitude (x-axis) correspond to geometric height variations $<1R_\oplus$ (y-axis). Instead, the location of the photosphere (scatter points) is more sensitive to the metallicity (line colours) -- which we explore as a grid parameter (Table~\ref{tab:params}) -- and measurement wavelength  (marker shapes). Wavelength sensitivities arise due to the fundamentally spectroscopic nature of gaseous absorption \citep{pierrehumbert_book_2010, stamnes_radiative_2017}, although our main results remain agnostic to effect, allowing generalised comparison against observations made by diversity remote-sensing facilities \citep{yurchenko_data_2025, madhusudhan_exoplanetary_2019}.

Our incorporation of 1D thermochemical calculations into whole-planet structure retrievals represents an advancement beyond previous planet-structure modelling. Future developments could attempt unification of whole-planet structural retrievals directly with spectroscopic measurements of photospheric compositions. However, this would require careful consideration and modelling of  disequilibrium processes known to act upon these low-pressure regions \citep{madhusudhan_exoplanetary_2016}. For example,  ultraviolet-driven photochemistry \citep{tsai_photochemically_2023, nicholls_temperaturechemistry_2023} and ionisation processes \citep{helling_exoplanet_2023,munoz_heating_2023}.

%%%%%%%%%%%%%%%%%%%%%%%%%%%%%%%%%%%%%%%%%%%%%%%%%%

% Don't change these lines
\bsp	% typesetting comment
\label{lastpage}
\end{document}